\documentclass[usenatbib,useAMS]{mn2e}

\usepackage{longtable}
\usepackage{times}
\usepackage{epsf}
\usepackage[dvips]{epsfig}

\begin{document}

\label{firstpage}

\title[Formation of dSph galaxies]{A Possible Formation Scenario for
  Dwarf Spheroidal Galaxies - II: A Parameter Study}
\author[Assmann et al.]{
  P. Assmann$^{1,2}$ \thanks{E-mail: passmann@astro-udec.cl},
  M. Fellhauer$^{1}$ \thanks{mfellhauer@astro-udec.cl} ,
  M. I. Wilkinson$^{3}$ \thanks{miw6@astro.le.ac.uk},
  R. Smith$^{1}$ \thanks{rsmith@astro-udec.cl}
  M. Bla\~{n}a$^{1}$ \thanks{mblana@astro-udec.cl}\\
  $^{1}$ Departamento de Astronom\'{i}a, Universidad de
  Concepci\'{o}n, Casilla 160-C, Concepci\'{o}n, Chile \\
  $^{2}$  Departamento de Astronom\'{i}a, Universidad de Chile, Camino
  El Observatorio 1515, Las Condes, Santiago, Chile \\
  $^{3}$ Department of Physics \& Astronomy, University of Leicester,
  University Road, Leicester LE1 7RH, UK}

\pagerange{\pageref{firstpage}--\pageref{lastpage}}
\pubyear{2011}

\maketitle

\begin{abstract}
Dwarf spheroidal (dSph) galaxies are considered the basic building
blocks of the galaxy formation process in the $\Lambda$CDM (Lambda
Cold Dark Matter) hierarchical cosmological model.  These galaxies are 
believed to be the most dark matter (DM) dominated systems known, have
the lowest stellar content, and are poor in gas.  Many theories
attempt to explain the formation of dSph galaxies resorting to the
fact that these galaxies are mainly found orbiting large galaxies or
invoking other mechanisms of interactions.  Here we show the full set
of simulation as an extension of our fiducial model, where we study the
formation of classical dSph galaxies in isolation by dissolving star
clusters within the DM halo of the dwarf galaxy.  In our parameter
survey we adopt cored and cusped DM halo profiles and consider
different numbers of dissolving star clusters.  We investigate the
dependency of observable quantities with different masses and
scale-lengths of the DM halo and different star formation efficiencies
(SFE).  We find that our proposed scenario explains many features of
the classical dSph galaxies of the Milky Way, like their morphology
and their dynamics.  We see trends how the surface brightness and the
scale-length of the luminous component vary with the parameters of our
simulations.  We also identify how irregularities in their shape,
i.e.\ clumpiness and ellipticity vary in our simulations.  In velocity
space, we identify the parameters leading to flat velocity dispersions
curves.  We recognize kinematically cold substructures in velocity
space, named fossil remnants and stemming from our unique initial
conditions, which alter the expected results.  These streaming
motions are considered as a key feature for future observation with
high resolution to validate our scenario. 
\end{abstract}


\section{Introduction}
\label{sec:intro}

Dwarf spheroidal galaxies (dSph) have sparked great interest in
astronomy in the latest years, since they would be the seed of the 
formation of larger structures in the cold dark matter cosmology.
They are believed to be highly dark matter dominated objects.   The
observed dSph galaxies are characterized by low surface brightnesses
with an absolute magnitude ranging between $-13 \leq$ M${_V}$ $\leq-9$
\citep{mat98, bel07}, half-light radii from $40$ to $1000$~pc
\citep{mat98, sim07, mar08}, low central concentrations and old
stellar populations in almost all cases.   Their mass in stars is low
(less than $10^8$M$_{\odot}$) and~\citep[with the exception of  Leo~T, see][]{RW08,irw07} they are devoid of HI gas \citep[][]{grc09}.
However, due to their intrinsic faintness the study of dsph galaxies has been very difficult.

During the past decade, in the era of SDSS (Sloan Digital Sky Survey)
and large spectroscopic surveys with 8-meter-class telescopes, our
understanding about this objects in our Local Group has grown
substantially.  New dSph galaxies have been found, doubling the sample
of the nine dSphs that we already knew \citep{wil05, bel06, zuc06,
  bel07, sak06, irw07, wal07}.  Furthermore, several known dSphs
galaxies (e.g.\ Fornax, Sculptor, Sextans, Ursa Minor) show signs of
stellar substructure or multiple distinct chemo-kinematic populations
\citep{col04, col05, mac03, wal06, bel01,bat08, bat11,kle04}.
Specifically, in Fornax it is observed, that there are stellar
over-densities along the minor axis and five globular clusters,
possibly remnants of past mergers \citep{col04, col05, mac03} or an
unknown formation mechanism.  Although, enormous advances in measuring
and understanding the properties of these faint objects are made,
there is not a definitive model of a possible formation scenario, that
can predict all the known properties.

There are many models that attempt to explain the possible formation
of these dSphs \citep{kor12}.  One class considers the dark matter cosmology as a
starting point, assuming dwarf disc galaxies embedded in DM haloes as 
initial models. To convert these objects into the dSph galaxies we
know, external processes have to be invoked.  These external
influences could stem from a larger galaxy (e.g.\ like the MW),
exerting tidal and ram pressure stripping forces (e.g. \citet{ein74,
  fab83, gne99, may01, kra04, may06, may07, kli09, kaz11,pop12}) onto the
dwarf.  In these models, the morphological transformation by tidal
forces of late-type dwarfs, leading to objects resembling present-day
dIrr, combined with mass loss due to tidal and ram pressure stripping
aided by heating due to the cosmic ionized background can turn the
dwarfs into objects with low angular momentum, high mass-to-light
ratios, faint stellar luminosity profiles, and velocity dispersion
profiles that resemble classical dSph galaxies.  These models have
difficulties to explain the presence of distant isolated dSph galaxies
such as Tucana and Cetus.  
 
Another mechanism considers resonant stripping \citep[introduced
by][]{don09}. The models start with the same initial conditions as
described above, i.e.\ they start with rotationally supported
galaxies, but this time consider resonant stripping by encounters
between dwarf disc galaxies.  The dSph galaxies are formed by a
process driven by gravitational resonances.   

There is another model that attempts to explain the formation of these
dwarf galaxies, without assuming DM haloes at all, i.e.\ leaving the
standard cosmological framework.  The model is based on energy and
momentum conservation, when gas-rich galaxies interact \citep{met07}.
In this model the dSph galaxies are regarded as second-generation
objects, devoid of DM, forming in the tidal tails of gas-rich
interacting galaxies.  

All the models cited above consider the interaction between two or
more galaxies to explain the formation of dSph galaxies.  Here, we
present a different approach, considering observational evidences for
kinematic substructures \citep{wil02}.  These properties are
consistent with the kinematics of disrupted star clusters, i.e.\
stellar streams.  

We investigate, numerically, initial conditions for isolated models
that allow for the formation of objects that resemble the classical
dSph galaxies.  Our model is based on the assumption that stars never
form in isolation but in a clustered way like associations and star
clusters \citep{tut78}.  The dynamical evolution of these star clusters, i.e.\ their
dissolution due to gas expulsion and subsequent formation of a diffuse
stellar distribution in the centre of a DM halo, may explain the
formation of classical dSph galaxies, including all their
irregularities in the stellar and kinematic distribution as well as
surviving star clusters around them.  We envisage that our scenario could be immediately
relevant for forming dSphs from initial conditions similar to Leo~T which has low luminosity, little evidence of angular momentum 
in its gas component~\citep{RW08}, and show no signs of being perturbed by the tidal forces from
nearby galaxies (i.e. the Milky Way in the case of Leo~T.)

In our first paper \citep{ass12} we introduced our scenario, showed
and discussed the results of a fiducial model, which matches the
observations of classical dSph galaxies.  We also presented possible
ways to verify our scenario.  In this paper now, we show how our
results depend on the initial conditions of our simulations by
performing a wide parameter search.


\section{Setup}
\label{sec:setup}

\subsection{Dark matter haloes}
\label{sec:dmhaloes}

One of the parameters we choose for testing our formation scenario
is the profile of the DM halo.  We consider two profiles for the dark matter - cusped and cored. 
Standard, dark matter only cosmological simulations suggest that galaxy haloes should have cusped inner profiles~\citep[see e.g.][; hereinafter NFW]{nav97}. We generate initial conditions for our cusped NFW haloes using NEMO, which is
described in \citet{deh05}.  On the other hand, we note that dynamical models of several Local Group
dwarf galaxies suggest that these galaxies may have DM haloes with cored profiles~\citep{kle03,gil07,goe06, col12}. Further, there is also evidence from the stellar kinematics for the presence of cored profiles in these objects \citep{bat08,wal11}. Therefore, we also consider models with a cored DM halo
using a Plummer profile \citep{plu11}.  Both types of DM halo
profiles are modeled using $1,000,000$ particles.

We adopt two variable parameters related with the DM halo.  The
first one is the mass enclosed within $500$~pc, M$_{500}$.  The second 
is the scale-length R$_{\rm h}$ of the DM halo.  We consider masses
M$_{500} = 10^{7}$~M$_{\odot}$, $4 \times 10^{7}$~M$_{\odot}$, and
$10^{8}$~M$_{\odot}$.  For each mass we further investigate three
different scale-lengths for the DM halo.  We adopt R$_{\rm h}$ equal
to $0.25$, $0.5$, and $1.0$~kpc.

The corresponding virial radii $R_{\rm vir}$ and concentration
parameters $c = R_{\rm vir}/R_{\rm h}$ for the NFW haloes can be found
in Tab.~\ref{tab:init-cond}. 

\subsection{Luminous component}
\label{sec:scluster}

The luminous components of our simulations are formed by the evolution
of many ($N_{0}$) star clusters within the DM halo.  We model each star
cluster as a Plummer sphere \citep{plu11}, following the recipe of
\citet{aar74}.  The Plummer radius, R$_{\rm pl}$, is $4$~pc and the
cut-off radius, R$_{\rm cut}$, is $25$~pc.  The values for R$_{\rm
  pl}$ and R$_{\rm cut}$ are based on the observations of young star
clusters in the Antennae galaxy \citep{whi99}.  All star clusters are
represented by 100,000 particles each.   

There are many processes acting on star clusters, for example
gas-expulsion, stellar evolution, relaxation, external tidal fields.
However, gas-expulsion can be considered the most important.  From the
observations in Antennae galaxy \citet{fal06} found that the number of
clusters drop exponentially with age and that the median age of
clusters is only $10^{7}$~years.  In addition, in the Milky Way
\citet{lad03} speculate that only a small fraction of embedded
clusters ($~10\%$) can become open clusters.  Motivated by this facts,
we add the process of gas-expulsion to our model and we study how the
results are affected considering different values of SFE for the star
clusters. 

We implement this process in the following way: 
\begin{itemize}
\item We assume a final total luminous mass of the dSph galaxy of
  $4.5\times10^{5}$~M$_{\odot}$.  These value lies in the range of
  stellar masses of the classical dwarfs \citep{mat98}.
\item This mass is now divided into $N_{0}$ star clusters giving a
  final mass of a single cluster.  For simplicity we assume that all
  star clusters have the same mass and do not take a mass spectrum
  into account.
\item Now we adopt a SFE and alter (i.e.\ increase) the mass of a
  single cluster according to its mass in the embedded phase, i.e.\
  mass of stars and gas.  In this form we set up the embedded cluster
  as a Plummer sphere in virial equilibrium.
\item After starting the simulation we mimic the gas-expulsion by
  reducing the mass of the star cluster during one of its
  crossing-times back to the mass of the stars alone, leaving the
  cluster out of virial equilibrium and slowly dissolving.  
\end{itemize}

In the simulations we vary the initial number of star clusters within
the DM halo and the SFE.  We consider an initial number, $N_{0}$, of
15, 30 and 60 star clusters.  These numbers have been chosen
arbitrarily. 

The star formation rate and efficiency in a dSph galaxy is supposedly
low \citep{bre10}.  Therefore, we adopt values of SFE of $15$ and $30$
per cent for each star cluster.  But we also check if a high value of
the SFE, i.e.\ $60$ per cent, allows the formation of an object that
resembles a dSph galaxy.  

The star clusters are placed inside the DM halo following a Plummer
distribution.  We choose this distribution, taking into account the
fact that we expect more star clusters to form in the central area
than further out in the dwarf.  The initial number of clusters is that
low, that the exact form of the distribution is not of importance, as
different distributions could not be differentiated due to low number
statistics.  Initially, these star clusters are in virial equilibrium
on their orbits inside the DM halo.  The initial orbital velocities
are obtained using the Jeans equation \citep[see][]{bin87}: 
\begin{eqnarray}
  \label{eq:jeans}
  \sigma^{2}_{r,i}(r) & = &
  \frac{1}{\rho_{i}(r)}\int^{rc}_{r}\frac{GM_{tot}(r')}{r'^{2}}
  \rho_{i}(r')dr',
\end{eqnarray} 
where M$_{\rm tot}$ corresponds to the total mass given by the sum of
the mass of all embedded clusters and the mass of the DM halo in each
simulation. 

As the mass of the halo is much higher than the mass of the expelled
gas, we do not consider the lost gas-mass in our simulations, apart
from its effect on the SCs, which is leaving them out of virial
equilibrium and dissolving. 

The distribution of the star clusters has a scale-length, $R_{\rm
  sc}$, which we vary to be $250$, $500$, and $1000$~pc.

\begin{table}
\caption{Table explaining the labels of all our simulations.  The
    first column specifies what type of DM halo profile we are using.
    N is used to refer to a NFW profile and P to a Plummer profile.
    The second column indicates the number of SCs at the beginning of
    each simulation, where 015, 030 and 060 correspond to 15, 30 and
    60 star clusters, respectively. The next two columns are used to
    denote the mass enclosed at the radius of $500$~pc and the
    scale-length of the DM halo, respectively. For example, M 407
    is used to denote that M$_{500}=4.0 \times 10^{7}$~M$_{\odot}$ and
    RH 025 that R$_{h}=0.25$~pc.  Similarly, the fifth column denotes
    the scale-length of the initial distribution of the star clusters.
    In the sixth column we indicate the SFE: 15\%, 30\% and 60\%.} 
    \centering
  \label{tab:conf}
  \begin{tabular}{cccccc}\\ \\ \hline
    N/P & $N_{0}$ & $M_{500}$ & $R_{\rm h}$ & $R_{\rm sc}$ & SFE \\
    \hline
    N & $015$ & M  $107$ & RH $025$ &  RS  $025$ & S$15$ \\
    P & $030$ & M  $407$ & RH $050$ &  RS  $050$ & S$30$ \\
      & $060$ & M  $108$ & RH $100$ &  RS  $100$ & S$60$  \\
    \hline
  \end{tabular}
  \end{table}

In Table~\ref{tab:conf}, we show the nomenclature that we use to label
each simulation.  The first column specifies what type of DM halo
profile we are using.  N is used to refer to a NFW profile and P to a
Plummer profile.  The second column indicates the number of SCs at the
beginning of each simulation, where 015, 030 and 060 correspond to 15,
30 and 60 star clusters, respectively.  The next two columns are used
to denote the mass enclosed at the radius of $500$~pc and the
scale-length of the DM halo, respectively.  For an example, M 407 is
used to denote that M$_{500} = 4.0 \times 10^{7}$~M$_{\odot}$ and RH 025 
that R$_{h} = 0.25$~pc.  Similarly, the fifth column denotes the scale
length of the initial distribution of the star clusters, $R_{\rm sc}$.
In the sixth column we indicate the SFE: 15\%, 30\% and 60\%. 

A more detailed view of the initial conditions of all our simulations
can be found in Tab.~\ref{tab:init-cond}.

\subsection{The simulations}
\label{sec:code}

As described above we start our simulations with all components in
virial equilibrium.  All star clusters are inserted at the same time
for simplicity.  This does not imply that our models depend on the
fact that a dSph galaxy has just a single star formation episode.  It
just reduces the number of free parameters.  We leave the
investigation of more complicated forms of star formation histories to
a follow up project.

Within the first internal crossing time of the star clusters we remove
their assumed gas-mass by artificially reducing the mass of each
particle.  This leaves the cluster out of equilibrium, expanding and
slowly dissolving along its orbit inside the DM halo.

The simulations are performed using the particle-mesh code {\sc
  Superbox} \citep{fel00}, which has additional two levels of
high-resolution grids for each modeled object.  Those high resolution
grids are moving with the objects through the simulation space
(focused on the centre of density of an object).  The resolution of
the highest resolution grid has usually $67$~pc per cell for the DM
halo and $0.8$~pc per cell for the star clusters.  This grid-level
monitors the central area of the DM halo and the star clusters
respectively.  The mid-level grids have resolutions of $333$~pc per
cell for the halo and covers the complete area in which the star
clusters are moving and $166.6$~pc for the star clusters to resolve
the interactions between the dissolving clusters properly.  The
outermost grid has the same size for all modeled objects and covers
the simulation area beyond the virial radius of the DM halo, with a
resolution for most simulations of about $1.6$~kpc per cell.  At a
later stage of the simulation, when all star clusters are dissolved
and have formed a fuzzy luminous object, we re-order the stars into
one luminous object with grid-sizes for the high resolution grids
depending on the size of the formed object.  This is done for
resolution and performance reasons, likewise.

The code uses a fixed time-step of $0.25$~Myr to ensure that the
internal dynamics of the initial clusters are resolved properly.  We
follow the evolution of our models over a long period of time, i.e.\
$10$~Gyr and give the results of our models at the end of this time
period.  We study the final object by analyzing its morphology,
luminosity and dynamics. 

One might argue that a particle-mesh code is not suitable to perform
simulations of star clusters as it is by definition collision-less and
neglects internal two-body processes of the star clusters, which are
important for the long-term evolution of the cluster.  But as almost
all clusters in our simulations dissolve and spread their stars
throughout the central region of the dwarf galaxy quite fast and we do
not investigate the properties of surviving clusters, the stars in
our simulations enter a collision-less state very quickly and are very
well modeled by a collision-less code.  As a final remark, we state
that the star-particles in our simulations do not represent single
stars at all.  The particles of a mesh-based code rather represent
tracers of the phase-space.  Therefore the actual number of
star-particles does not necessarily coincide with the number of
stars.  More information can be found in \citet{fel00}.


\section{Results}
\label{sec:res}

We perform $73$ numerical simulations of which $53$ simulations have
different initial parameters.  The $20$ other simulations are
repetitions of the same parameter sets with different random number
seeds to analyze the importance of the low number statistics regarding
the initial positions and velocities of our $N_{0}$ star clusters.
The goal is to show that our formation scenario of dSph galaxies leads
to objects that resemble these galaxies, even when quite different
initial conditions are assumed thus demonstrating the plausibility of
our scenario.  

\subsection{Surface density profile}
\label{sec:surface}

\begin{figure*}
  \epsfxsize=16cm
  \epsfysize=16cm
  \epsffile{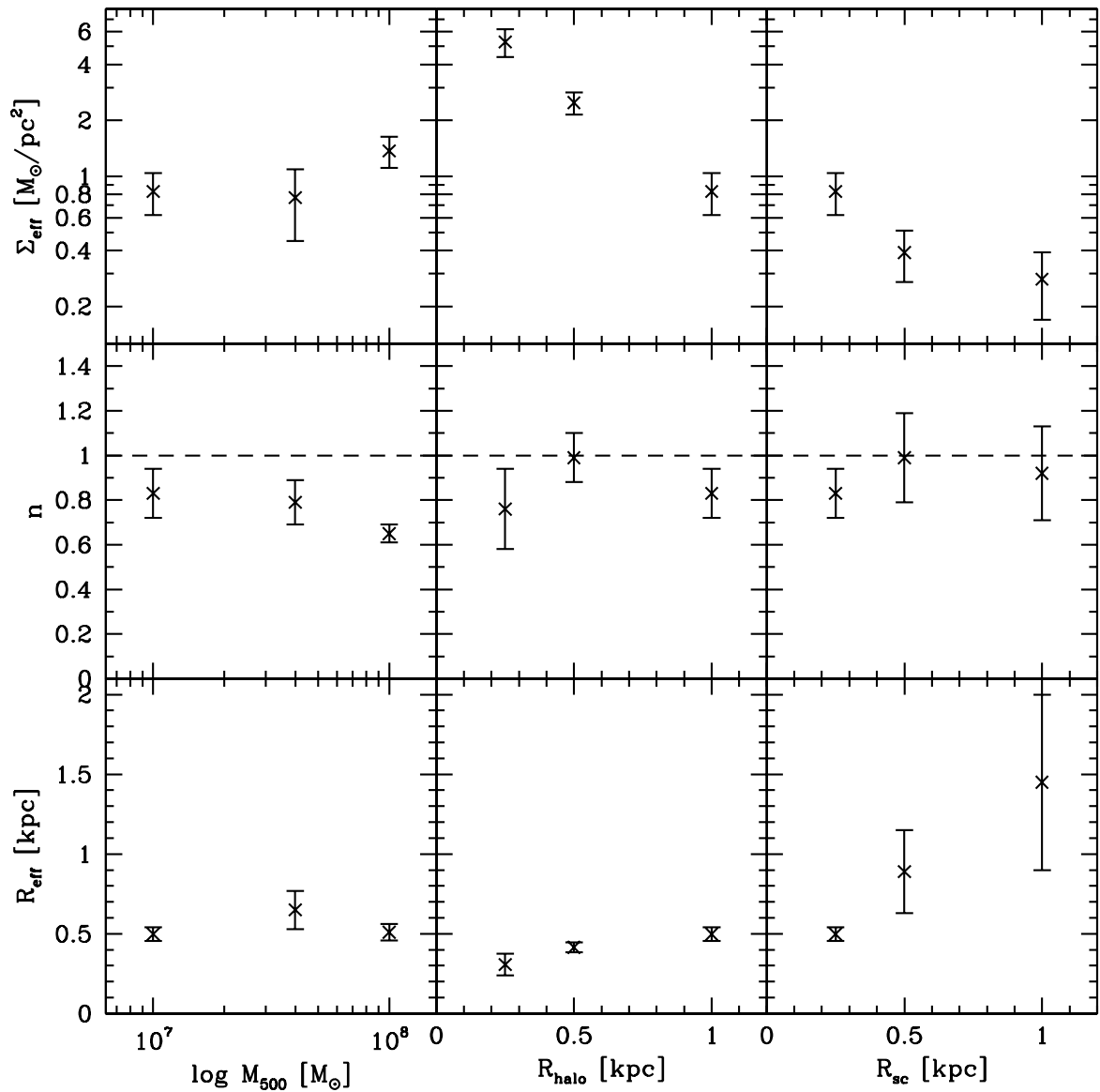}
  \centering
  \caption{Correlation between the S\'ersic parameters of our objects 
    and the parameters of the initial conditions: $M_{500}$, $R_{\rm h}$
    and $R_{\rm sc}$ keeping all other parameters constant.  In the
    first column we show the dependency of the S\'ersic parameters as 
    function of the mass of the halo within $500$~pc.  In these
    simulations we keep all the other parameters constant, i.e.\
    $R_{\rm halo} = 1.0$~kpc and $R_{\rm sc} = 0.25$~kpc.  In the
    middle column we show simulations whose $M_{500}= 1 \times
    10^{7}$~M$_{\odot}$ and R$_{sc}=0.25$~kpc are kept constant and we
    vary the scale-length of the halo $R_{\rm h}$.  In the right
    column we consider simulations with $M_{500}= 1 \times
    10^{7}$~M$_{\odot}$ and $R_{\rm h}=1.0$~kpc and vary the
    scale-length of the star cluster distribution $R_{\rm sc}$. The
    dashed line in the middle row shows $n = 1$, which corresponds to
    an exponential profile, as seen with spheroidal galaxies.} 
  \label{fig:bright}
\end{figure*}

Several surface density profiles have been used in the literature to
fit the observed surface brightness curves of the dSph galaxies
\citep{irw95,maj05}. There is also some evidence of environmental effects in the surface brightness profiles of dSphs \citep{bel96,mac06,rea06} .  In this work we consider S\'ersic \citep{cao93}
profiles according to the following formula:
\begin{eqnarray}
  \label{eq:sersic}
  \Sigma(R) & = & \Sigma_{\rm eff} \exp \left( -b_{n} \left[ \left(
        \frac{R}{R_{\rm eff}} \right)^{1/n} - 1 \right] \right) \\
  b_{n} & = & 1.9992 n - 0.3271 
\end{eqnarray}
where R$_{\rm eff}$ is the effective radius and $\Sigma_{\rm eff}$ is the
surface density at the effective radius.  The index $n$ gives
information about the shape of the dSph galaxy.  In the case $n
\approx 1$, we have a exponential-shaped profile for the
surface density distribution. 

Surface brightness profiles show integrated light along the
line-of-sight.  As we do not know along which line-of-sight our
objects may be observed, we calculate a mean value out of three
possible sight-lines along the Cartesian coordinate axes $x$, $y$ and 
$z$.  Some of the simulations have been repeated using a different
random number seed to explore the variations of the properties of the
objects formed from different realizations of the same initial
conditions.   

The idea for considering S\'ersic profiles comes from the fact that
King profiles, even though widely used, do not 
reflect that we are analyzing luminous components inside DM haloes,
i.e.\ well shielded from any external potential.  Therefore, the
objects can not be expected to be tidally truncated.  Whereas some of
the known dSph galaxies seem to be well fitted by King profiles, there
is no physical reason behind it.  We find that many of our models are
not well fitted by King profiles with a truncation radius, 
i.e.\ a tidal radius.  As our initial conditions for the stars are
based on Plummer distributions it is quite natural to try as well a
Plummer fit to the results.  Again, our results show that the 
luminous components are badly fitted by two-parameter Plummer
profiles.  This result can be seen as a confirmation that our choice
of the initial distribution of the luminous mass does not affect the
final outcome.  Finally, the S\'ersic profile has the advantage of
being a 3-parameter fit that easily adjusts to the data.  A detailed
overview of all profiles fitted to all the simulations can be found in
Tab.~\ref{tab:res1}. 

In Fig.~\ref{fig:bright}, we show the correlation between the three  
parameters of the S\'ersic profile ($\Sigma_{\rm eff}$, $n$ and
$R_{\rm eff}$) fitted to our models with the parameters M$_{500}$,
R$_{\rm h}$ and R$_{\rm sc}$ of our initial conditions, while keeping
the other two parameters constant.  By analyzing the data
we do not find any dependency of the S\'ersic parameters on the halo
profile used (Plummer or NFW) or on the number of star clusters
$N_{0}$.  Also, as long as the SFE is low (i.e.\ 15 or 30~\%), there
is no dependency of the results either.  We therefore calculate the
mean values of the fitting parameters keeping the above mentioned
parameters constant but adding up all simulations with different
profiles, number of star clusters and SFE.  The data points show the
mean values and the error-bars give the one sigma deviations of the
mean.  

In the left column we show the dependency of the S\'ersic parameters
on the mass of the halo within $500$~pc.  In these simulations we keep
$R_{\rm h} = 1.0$~kpc and $R_{\rm sc} = 0.25$~kpc constant.  We do not
see a change in the surface brightness profile parameters when
changing the mass of the DM halo.  One might expect more stars to
settle in a deeper potential, but this seems not to be the case.
Obviously it is not the absolute value of the potential influencing
the light distribution.  This seems to be supported by observations,
as we do not have large differences in surface brightness profiles
between the dSph galaxies. 

Instead we do see a strong dependency of the effective surface
brightness on the scale-length of the halo as seen in the top panel of
the middle column of Fig.~\ref{fig:bright}.  The surface density
increases as we go to smaller halo scale-lengths.  This can be
understood as we have a steeper profile with smaller scale-lengths.
This contradicts in a sense our finding that we see no trend between
cusped and cored profiles.  But as the NFW profiles have a cusp in
their centre, the Plummer profiles are steeper in the outer parts.  We
therefore conclude that it is rather a dependency on the steepness of
the halo-profile along all radii of the luminous component than just a
function of the very central behaviour.  Accordingly if we distribute
the mass more centrally concentrated, the scale-lengths of the
luminous components are slightly decreasing (bottom middle panel). 

Comparing our results with the classical dSph galaxies like Carina,
Draco or Sextans (whose total luminous mass agrees with our fixed
luminous mass), which have central surface-brightnesses in the order of
$1$--$3$~L$_{\odot}$\,pc$^{-2}$ \citep[][and references therein]{irw95},
our results favour larger halo scale-lengths of approximately
$0.5$~kpc and more (taking into account that our effective brightness
should be lower than the central one and that we expect a
mass-to-light ratio, for the stellar component only, of larger than
unity).  

A strong rising trend of $R_{\rm eff}$ with $R_{\rm sc}$ is observed
in the bottom-right panel of Fig.~\ref{fig:bright}.  This is also
expected because if the luminous matter started more concentrated, it
should also be more concentrated in the final object.  For R$_{\rm
  sc}=0.25$~kpc, the mean value of R$_{\rm eff}$ obtained in the
simulations is $\bar{R}_{\rm eff}= 498 \pm 44$~pc. For R$_{\rm
  sc}=0.5$~kpc  the resulting objects have an average effective radius
of $\bar{R}_{\rm eff}= 890 \pm 260$~pc.  Finally, for simulations with
an initial $R_{\rm sc}$ of $1$~kpc, the final effective radius has a
mean value of $1.45$~kpc.  Thus, the final scale-length is almost
double the initial scale-length of the star cluster distribution.
Here the dependency of the effective surface density is not that
pronounced but the dependency on the effective radius is quite strong.  

The results show that we can disentangle the influence of the
different initial parameters: 
\begin{itemize}
\item the mass of the halo has no influence on the mean values of the
  S\'ersic fits.
\item smaller scale-lengths of the haloes lead to brighter dwarf
  galaxies in the central area.  
\item if the luminous component is formed out of a smaller and denser
  gas-distribution (i.e.\ star cluster distribution) we see smaller
  scale-lengths in our final galaxy. 
\end{itemize}

Another reassuring result is shown in the panels of the middle row
of Fig.~\ref{fig:bright}.  We see that the S\'ersic index $n$ has no
dependency on any initial parameter.  It shows always mean values
around $1.0$, independently of the values of $M_{500}$, $R_{\rm h}$ or
$R_{\rm sc}$.  This means that our resulting objects have
approximately an exponential surface density profile, like the
dSph galaxies of the Local Group \citep{cao93,jer00,wal03}.

Photometrical observations of the dSph galaxies in the Local Group
\citep{irw95, mat98} show that we can find different sizes of these
galaxies.  For example, the effective radii of Draco, Ursa Minor,
Sculptor and Fornax are $180$, $200$, $110$ and $460$~pc,
respectively. 

If our formation scenario is correct, then the initial distribution of 
baryons which now form the luminous components had smaller
scale-lengths in their initial gas distribution.  As a rule by thumb
we can say: about half the value of their effective radii today.   

\subsection{Shape of the final objects in our simulations}
\label{sec:shape}

\begin{figure*}
  \centering
  \epsfxsize=16cm
  \epsfysize=16cm
  \epsffile{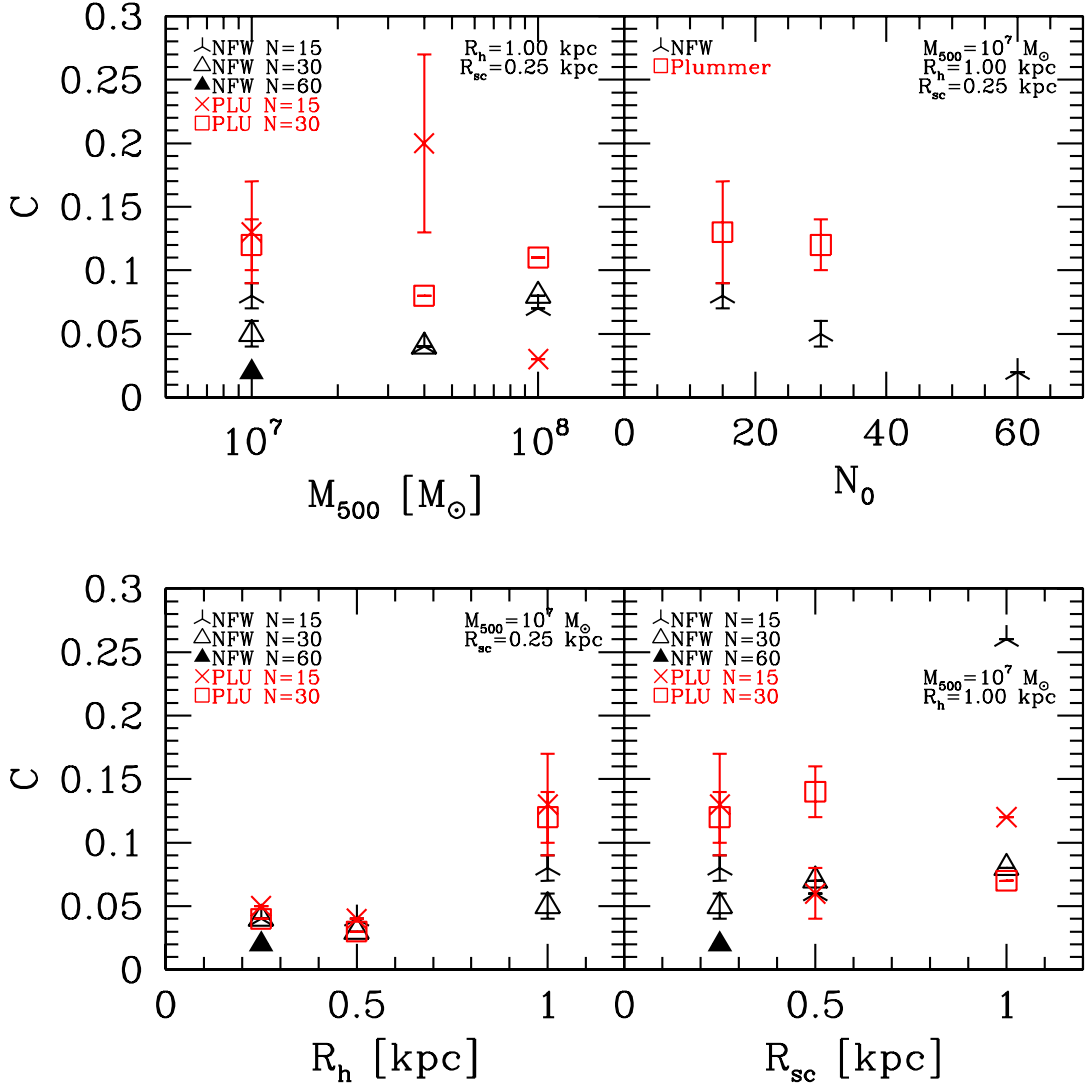}
  \label{fig:clumpshape1}
  \caption{Dependency of the shape parameter clumpiness $C$. In all
    panels (black) tri-shaped symbols (three-pointed stars and
    triangles) represent simulations with NFW haloes.  (Red) symbols
    with four corners (crosses and squares) belong to simulations 
    using Plummer haloes.  In all panels, except the top right, star
    symbols represent simulations using 15 star clusters, open symbols 
    represent 30 star clusters and the filled symbol in the top left
    shows the result of the simulation using 60 star clusters.  In
    the top left panel the dependency on $M_{500}$ is shown.  Here we
    keep $R_{\rm h} = 1.0$~kpc and $R_{\rm sc} = 0.25$~kpc constant.
    The top right shows the dependency on the number of star clusters
    $N_{0}$ keeping $M_{500} = 10^{7}$~M$_{\odot}$, $R_{\rm h} =
    1.0$~kpc and $R_{\rm sc} = 0.25$~kpc constant.  In the bottom left
    panel the dependency on the halo scale-length $R_{\rm h}$ is shown
    keeping $M_{500} = 10^{7}$~M$_{\odot}$ and $R_{\rm sc} = 0.25$~kpc
    constant.  And in the bottom right the dependency on the
    scale-length of the cluster distribution $R_{\rm sc}$ is shown,
    again keeping $M_{500} = 10^{7}$~M$_{\odot}$ and $R_{\rm h} =
    1.0$~kpc constant.  Symbols without error-bars are based on a
    single simulation only}
\end{figure*}

It is possible to characterize the shape of our resulting objects
considering the parameters of Clumpiness, $C$, introduced by
\citet{con03}, the ellipticity $e$ of the object, and the
isophotal parameter $A_{4}$ \citep{kho05}.  We show all results in
Tab.~\ref{tab:shape}.

\subsubsection{Clumpiness C}
\label{sec:c}

\begin{figure}
  \centering
  \epsfxsize=6cm
  \epsfysize=6cm
  \epsffile{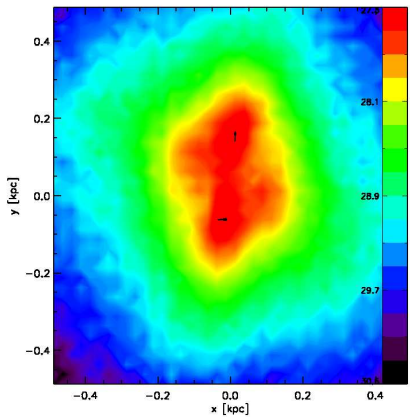}  
  \caption{Surface brightness contours of the simulation with the most
    extreme deviation from ellipticity, i.e. the highest clumpiness.}
  \label{fig:doublecore}
\end{figure}

One feature that can characterize the light distribution in a galaxy
is the patchiness of the distribution.  \citet{con03} introduce the
$C$-parameter (called clumpiness), which is a measure of the
inhomogeneity of the distribution of the stars.  We deviate from the
method explained in \citet{con03} and construct a smooth elliptic
model which fits our simulation results, using the IRAF routine {\sc
  Ellipse}.  Then we subtract this smooth model from our data and sum
the positive residuals.  The ratio between these residuals and the
original data is the clumpiness $C$.  More details about this method
can be found in \citet{ass12}.

In Fig.~\ref{fig:clumpshape1}, we show the relationship between the
clumpiness $C$ of the models and (top left) $M_{500}$, (top right)
$N_{0}$, (bottom left) $R_{\rm h}$  and (bottom right) $R_{\rm sc}$.
First we observe that all our simulation lead to low values of the
clumpiness parameter.  

It is hard to compare our results with the classical dSph of the MW as
there was not yet a determination of their clumpiness published.
Comparing our results with values published in \citet{con03} we see
that our results for cuspy halo profiles and small halo scale-lengths
lie in the range of values found for dwarf ellipticals and the results
for cored haloes and large scale-lengths at the lower end of the
values for dwarf irregulars.

Ten Gyr of evolution are enough to erase almost all deviations from a
smooth object in positional space.  That this is not necessarily true
for the velocity space will be dealt with in a later section. 

\begin{figure*}
  \centering
  \epsfxsize=16cm
  \epsfysize=16cm
  \epsffile{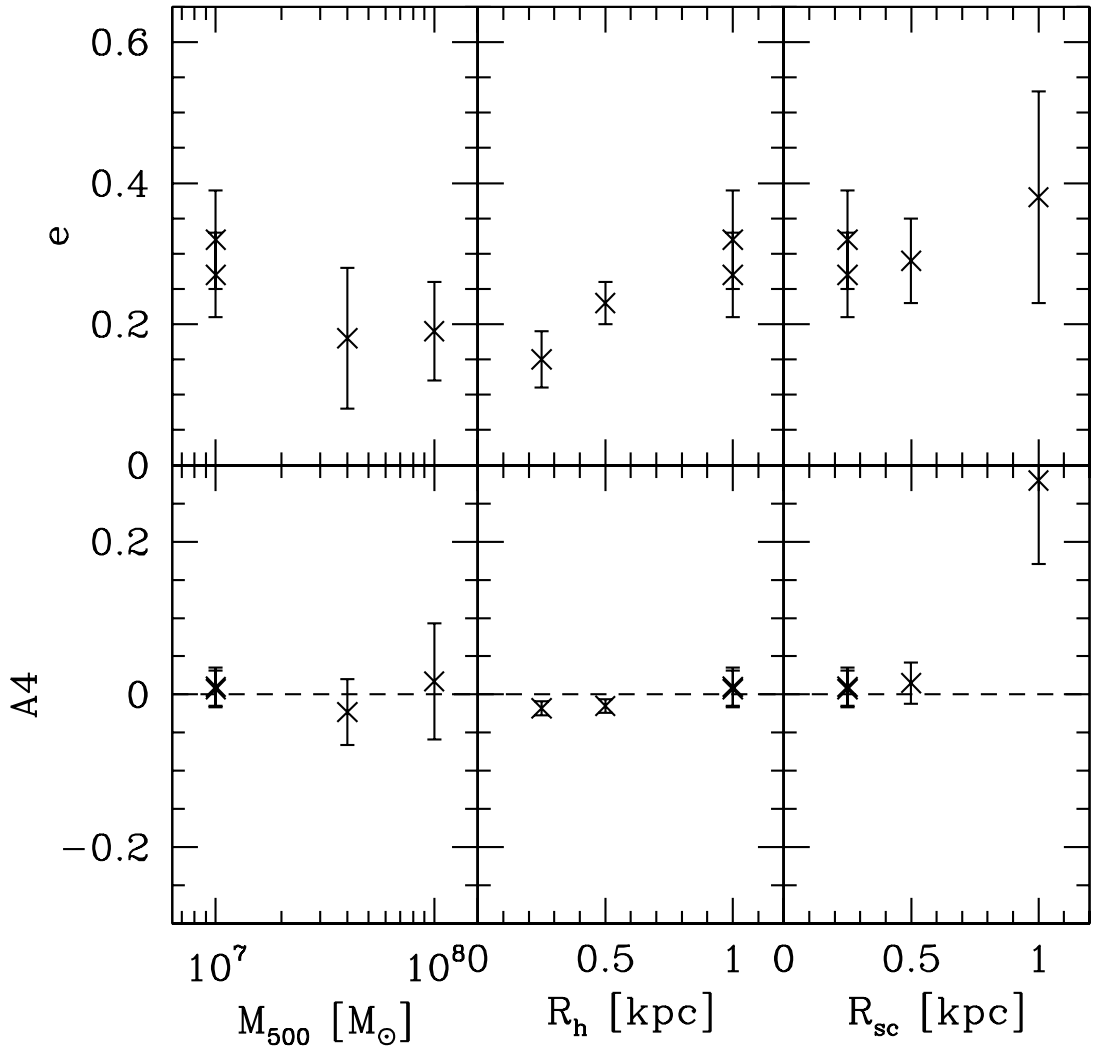}
  \caption{Ellipticity $e$ and isophotal deviation $A_{4}$ as function of
    the initial conditions $M_{500}$, $R_{\rm h}$ and $R_{\rm sc}$.
    In each panel we vary one parameter and keep the other two
    constant at the values $M_{500} = 10^{7}$~M$_{\odot}$, $R_{\rm h}
    = 1.0$~kpc and $R_{\rm sc} = 0.25$~kpc. The data-points with
    exactly those values are split between cusped and cored haloes to
    show that the halo shape has no influence (the cored profiles have
    slightly higher $e$ but well within the error-bars and the results
    for $A_{4}$ are indistinguishable).}
  \label{fig:ea4}
\end{figure*}

In all panels we observe that cored profiles (Plummer) lead to a
higher clumpiness than the corresponding cusped profiles (NFW).  This
in fact is not a new result and was expected.  Cuspy profiles
are better in erasing sub-structure.  Only if the scale-lengths of the
haloes are low, cuspy and cored profiles do not differ
significantly any longer (see bottom left panel of
Fig.~\ref{fig:clumpshape1}).  The mass of the halo (top left) and the
scale-length of the star cluster distribution (bottom right) seem to
have no influence on the clumpiness at all.  Finally we see a clear
trend with the number of star clusters (top right).  If we distribute
the mass into more star clusters the resulting clumpiness is lower.
This finding is expected as with an initial distribution which is 
'smoother' one ends up with a final object which has less deviations
from smoothness.

We can ask ourselves now, what does this 'clumpiness' mean for our
resulting objects.  We see a huge variety in shapes in our models,
like off-centre nuclei, secondary density peaks, peanut-shaped central
regions and as an extreme even two equally important density centres
as shown as an example in Fig.~\ref{fig:doublecore}.

\subsubsection{Ellipticity}
\label{sec:e}

Another structural parameter is the ellipticity, $e$.  This parameter
is defined by 
\begin{eqnarray}
\label{eq:ellip}
    e & = & 1 - \frac{b}{a},
\end{eqnarray}
where $b$ is the minor and $a$ is the major axis of an ellipse.  The
dSph galaxies of the Local Group Sculptor, Fornax, Ursa Minor, and
Draco have $e$ values of $0.32 \pm 0.03$, $0.31 \pm 0.03$,  
$0.56 \pm 0.05$ and $0.29 \pm 0.01$, respectively \citep{mat98}.

We determine the ellipticity values at the half-mass radius (as the
ellipticity may vary throughout the dwarf) by fitting a smooth model
to our simulation data using the IRAF routine {\sc Ellipse}. 

\begin{figure*}
  \centering
  \epsfxsize=16cm
  \epsfysize=16cm
  \epsffile{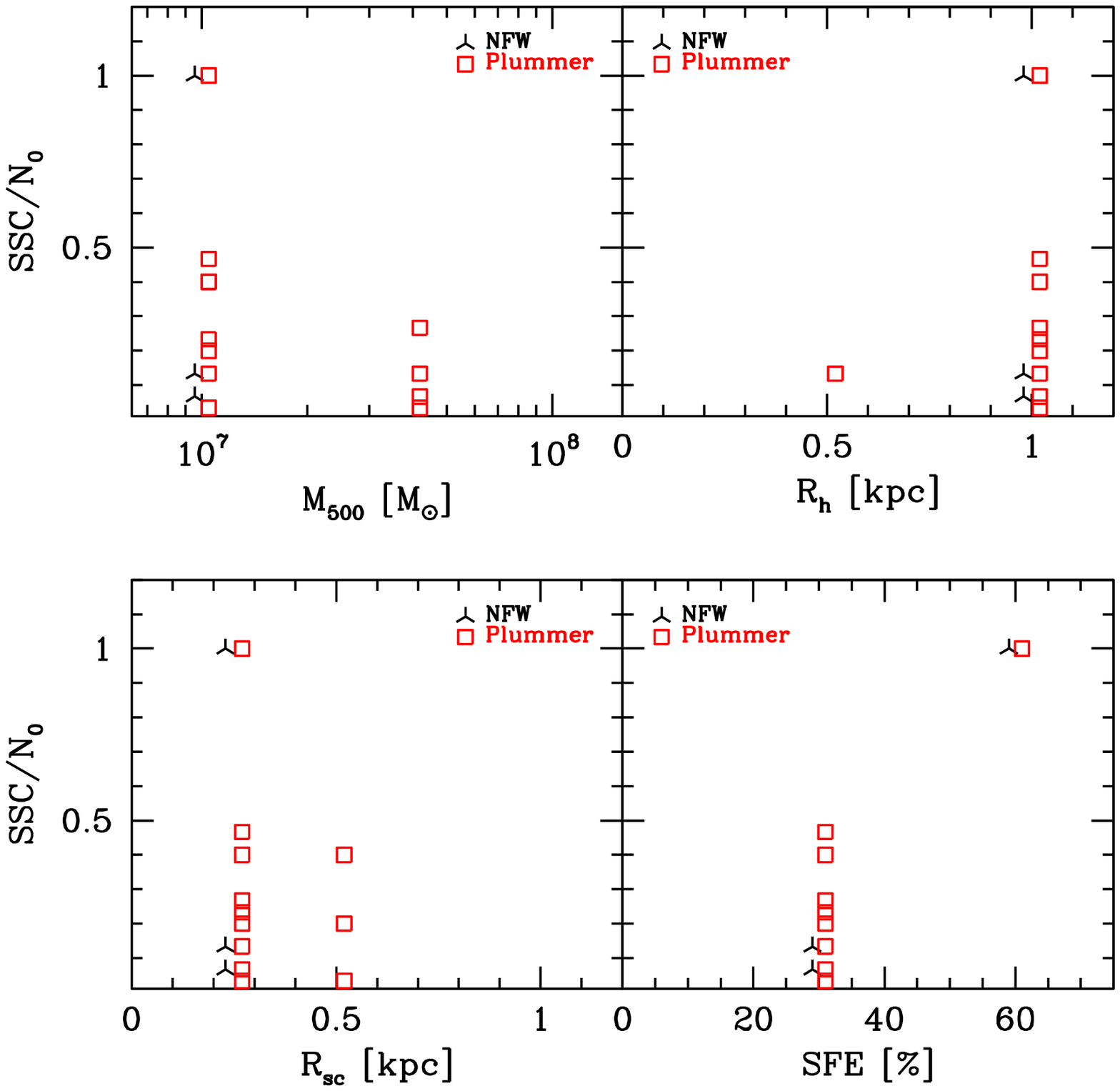}
  \caption{Surviving star clusters for all our simulations as function
  of the initial parameters $M_{500}$, $R_{\rm h}$, $R_{\rm sc}$ and
  SFE. (Black) tri-pointed stars are simulations with a cuspy halo,
  and (red) open squares show simulations with a cored halo-profile.}
  \label{fig:ssc}
\end{figure*}

In the top row of Fig.~\ref{fig:ea4}, we study the dependence of $e$
on the initial parameters of our numerical experiments by keeping all
other parameters constant.  We observe no dependency between $e$ and
M$_{500}$.  In the first bin ($10^{7}$~M$_{\odot}$) we show two data
points, for cusped and cored haloes separately, as we do have multiple
realisations of those parameter sets to determine mean values and
errors.  As they agree (within $1 \sigma$-error-bars) we do not
distinguish between the two halo types.  However, it is interesting to
observe that the final objects have $e$ values close to the values of
the classical dSph galaxies, regardless whether the simulations had a
cusped or cored DM halo \citep{mat98, mar08, mun10, san09, san10,
  mcg10}.   

It is also interesting that $e$ shows an increasing trend with the
scale-length of the DM halo.  At R$_{\rm h}=0.25$~kpc, $e = 0.15 \pm
0.04$.  When R$_{\rm h} = 0.5$~kpc $e = 0.23 \pm 0.03$.  Finally, for
R$_{\rm h} = 1.0$~kpc, we show again the two separate results from the
first panel and obtain $e = 0.27 \pm 0.06$ for NFW haloes and $e =
0.32 \pm 0.07$ for Plummer haloes.  A possible explanation could be
that with large scale-lengths more mass of the halo is at larger radii
and can influence radial orbits to be more eccentric, i.e.\ have
larger apo-centres.  If now, due to our low number in SCs, more orbits
are aligned in one direction leading to a non-spherical object, it
will be even more elongated if we have a larger scale-length of the
halo.   

If this trend should be real in spite the large error-bars, our data
would suggest that the real dwarf spheroidal galaxies of the MW would
have halo scale-lengths preferably of $0.5$~kpc or higher.  This
agrees well with our findings from the central surface-brightness.

There is no relation between the initial $R_{\rm sc}$ and the
ellipticity of the final object (see top right panel).  

\begin{figure*}
  \centering
  \epsfxsize=6cm
  \epsfysize=6cm
  \epsffile{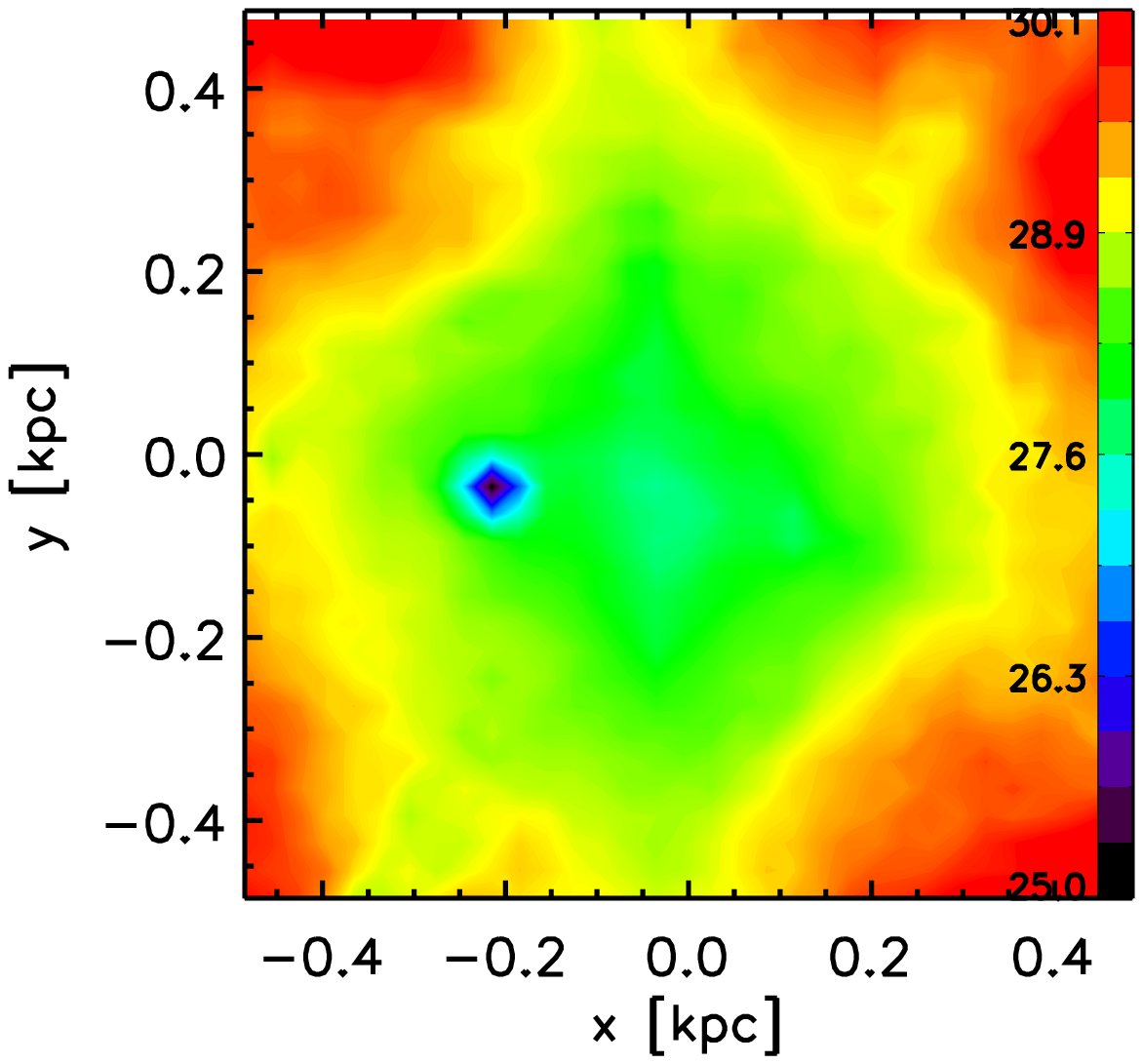}
  \epsfxsize=5cm
  \epsfysize=6cm
  \epsffile{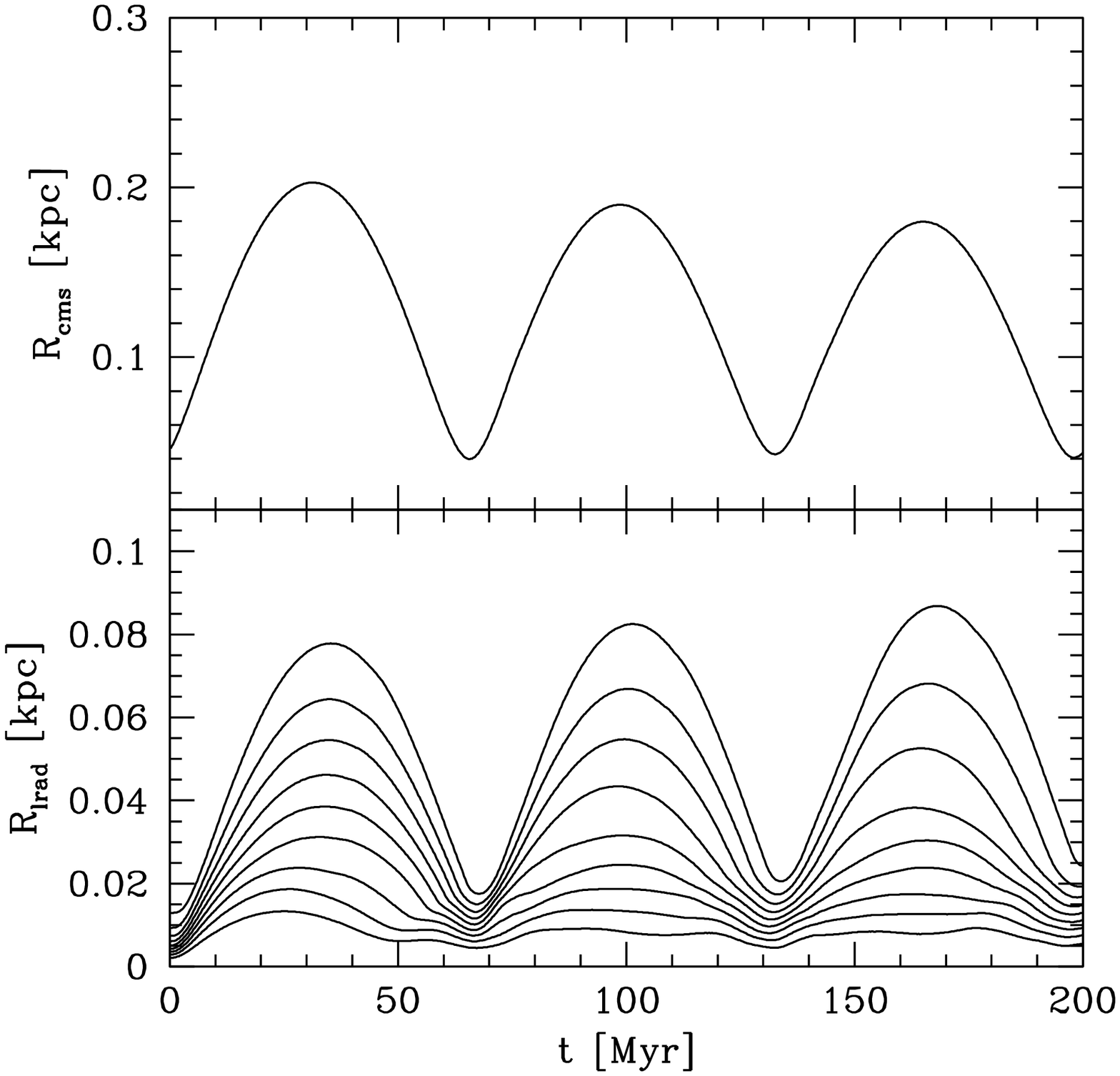}
  \epsfxsize=5cm
  \epsfysize=6cm
  \epsffile{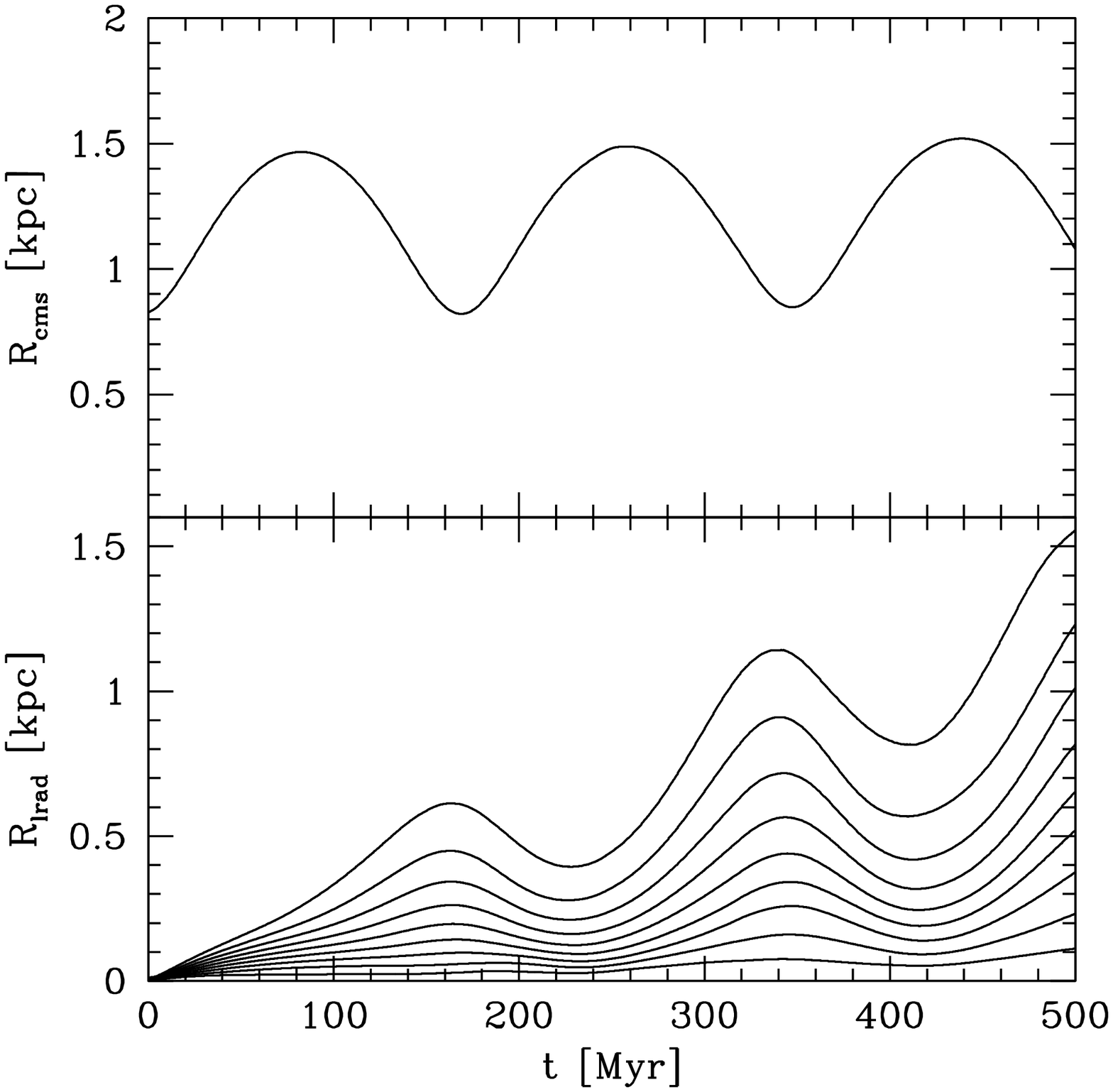}
  \caption{Why do star cluster survive?  Left: Surface brightness
    contours of a simulation showing a surviving SC in the central
    region.  Middle: The top graph shows the distance to the centre of
    the halo of the surviving SC, while the lower graph shows the
    evolution of the Lagrangian radii (10, 20,..., 90 per cent of the
    mass) with time.  Right: As a contrast we show the same data as in
    the middle panel for a dissolving SC.} 
  \label{fig:survive}
\end{figure*}

\subsubsection{$A_{4}$ parameter}
\label{sec:a4}

Another parameter obtained from the {\sc Ellipse} routine
is the fourth-order Fourier-coefficient, $A_{4}$, which corresponds to
the value of the isophotal deviation from a perfect ellipse.  $A_{4}$
is positive for 'disky' galaxies and negative for 'boxy' galaxies
\citep{kho05}.  The results of all our simulations are shown in
Tab.~\ref{tab:shape}. 

In the bottom row of Fig.~\ref{fig:ea4} we show $A_{4}$ versus
M$_{500}$, R$_{\rm h}$ and R$_{\rm sc}$.  As always, we keep the other
input parameter constant and only calculate the mean of simulations
with identical initial conditions but different random seeds.  In the
figure the parameter $A_{4}$, in general, tends to zero and this shows 
the spheroidal character of the objects.  There is no observable trend
between $A_{4}$ and $M_{500}$ or R$_{\rm h}$.  As we can see in all
panels (or better not see), is that the data-points for cusped and
cored haloes are almost identical and fall on top of each other.  For
$M_{500} = 10^{7}$~M$_{\odot}$, $R_{\rm h} = 1$~kpc and $R_{\rm sc} =
0.25$~ kpc we obtain for cusped haloes $A_{4} = 0.010 \pm 0.025$ and
for cored haloes $A_{4} = 0.007 \pm 0.024$.

We do not believe that there is an actual trend to 'disky' shapes with
high values of $R_{\rm sc}$.  This data-point is based on four
simulations only, from which three show 'extreme' deviations from 
ellipticity.  Such 'outliers' appear in every bin but usually with
alternating signs.  We think it is an effect by chance that for this
data-point all three 'outliers' have positive values.

The single values of the simulations are all in agreement with
observations of elliptical galaxies \citep{ben88}.  They all show only
small deviations in the order of 0.01 to 0.1 with alternating signs,
i.e.\ our formation process does favour neither disky nor boxy shapes.

\subsubsection{Surviving Star Clusters}
\label{sec:ssc}

We also indicate in Tab.~\ref{tab:shape} if we find star clusters,
which are not dissolved after the 10~Gyr of evolution in our
simulations.  We show these simulations also in Fig.~\ref{fig:ssc} and 
plot the relative number of surviving star clusters against the
initial conditions of the simulations.  We see in all
panels that in cored haloes, clusters have a higher chance to survive.
This suggests that traveling through a cuspy central region with
small scale-length (i.e. steep potential gradient) does not enhance
the survival-ability of the star clusters.  The bottom left panel
shows that more concentrated cluster distributions lead to more cluster
surviving. 

The bottom right panel shows the obvious.  If the SFE is as low as
$15$~per cent none of the cluster survives, independently how all
other parameters are distributed.  At a SFE of $30$~per cent we see
that in some simulations we have a few cluster surviving the
dissolution and destruction process, following the trends explained
above.  At a SFE of $60$~per cent we find the obvious result that none
of the star clusters gets disrupted.  They are bound enough to survive
their gas-expulsion and then the background potential is too strong,
i.e.\ the clusters have too high encounter velocities, such that the
star clusters can not merge.

The surviving star clusters are exclusively found in the very central
areas at $r < 200$~pc.  In the left panel of Fig.~\ref{fig:survive} we
show the surface brightness contours of a simulation, which hosts a
surviving SC.  The surviving SC is closely located to the centre.  The
simulation has a Plummer halo with a large scale-length of $1$~kpc.

In the middle panels we see the reason for its survival.  In the top
we show its orbit which has an apo-centre of only $200$~pc.  The lower
panel shows the Lagrangian radii of this cluster.  It expands due to
the gas expulsion during the first few Myr but the expansion is halted
and turned around after the SC reached its apo-centre and continued
orbiting towards the centre of the halo.  I.e.\ particles, which were
unbound and expanding get 'compressed' together again.  After the
peri-centre passage, which corresponds also with the time, the
particles have the smallest extension in space, the outer Lagrangian
radii are re-expanding.  With the $90$~per cent radius reaching about
$80$~pc at the apo-centre of $200$~pc, those mass-shells are in
reality unbound particles which follow the orbit of the original
cluster and are sometimes closer to peri-centre, together with the
remaining centre of density of the SC and sometimes they are more
distributed along the orbit at apo-centre.  But, the inner mass-shells
(10 to 20 per cent of the initial mass) do not expand again after the
first peri-centre passage.  They stay at an elevated but constant
level and form the surviving SC.

To contrast this findings we show in the right panels the evolution of
a dissolving SC.  In the top panel we see its orbit with per- and
apo-centre of $0.8$ and $1.5$~kpc respectively.  Again, we see an
initial expansion due to gas-expulsion in the first few Myr (in the
scale of the figure almost invisible). Then we see an oscillation in
the Lagrangian radii as well.  But, this oscillation happens on much
larger scales and only reflects the fact that we have orbital crowding
due to low orbital velocities of unbound particles at apo-centre -
therefore we have the minima in the Lagrangian radii close to the
apo-centres - and very long tidal tails close to peri-centre when the
orbital velocities of the particles are high - therefore we see the
maxima of the Lagrangian radii at or close to peri-centre. 

\subsection{Velocity Space}
\label{sec:vs}

\subsubsection{Velocity dispersion}
\label{sec:disp}

\begin{figure*}
  \centering
  \epsfxsize=16cm
  \epsfysize=16cm
  \epsffile{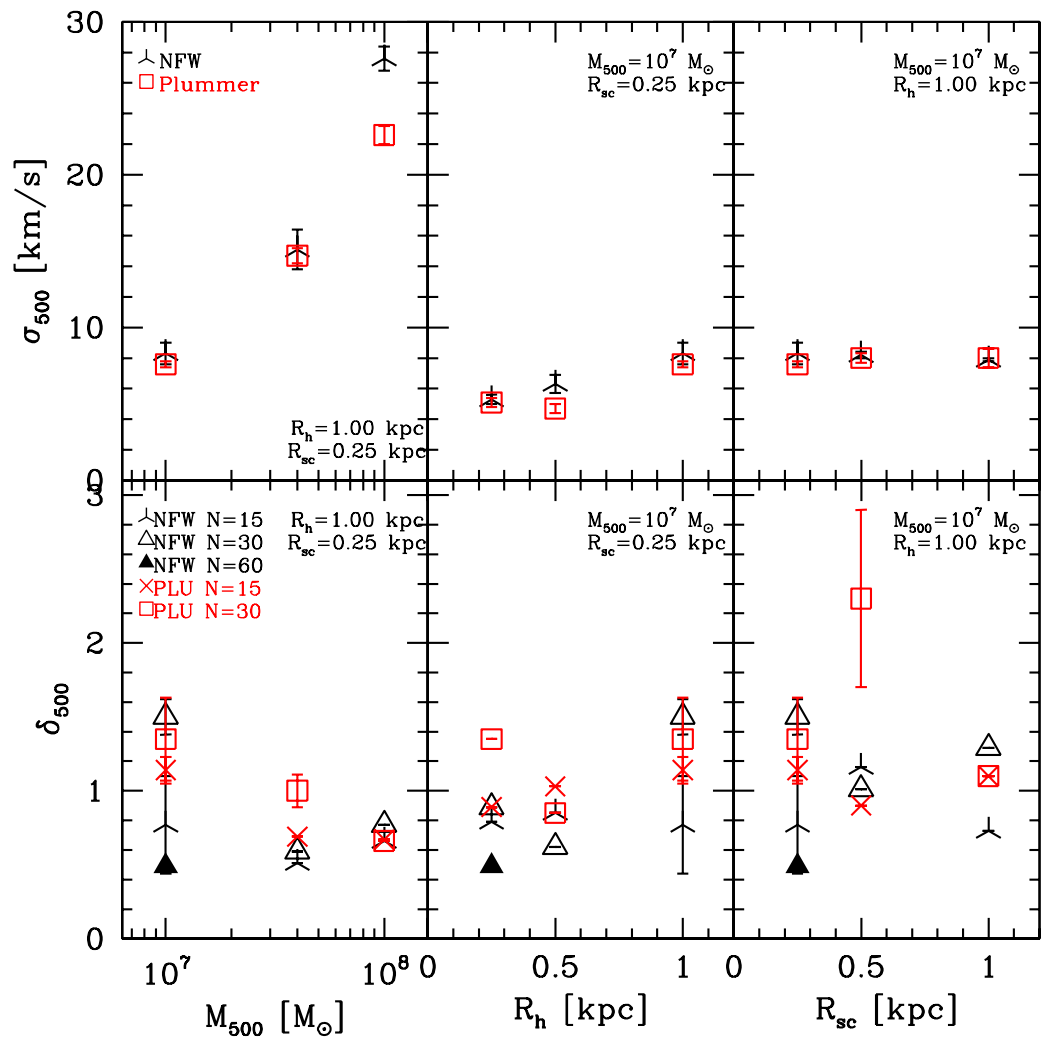}
  \caption{Line-of-sight velocity dispersion $\sigma_{500}$ (top
    row) and maximum deviation of the radial velocity
    $\delta_{500}$ (bottom row) as a function of the initial
    parameters of our models: $M_{500}$, $R_{\rm h}$ and $R_{\rm
      sc}$.  In the top row we show results for cuspy haloes as
    (black) tri-pointed stars and of cored haloes as (red) open
    squares.  In the bottom row we divide even further and show as
    (black) open triangles results of NFW haloes using $N_{0} = 15$
    star clusters (black) tri-pointed stars show the NFW haloes
    with $N_{0} = 30$ star clusters and the filled triangle is the
    result of the $N_{0} = 60$ simulation.  (Red) open squares are
    Plummer haloes with $N_{0} = 15$ and (red) crosses show Plummer
    haloes with $N_{0} = 30$.  As always we keep all other
    parameters constant as described in
    Fig.~\ref{fig:clumpshape1}.}
  \label{fig:sigma03bin}
\end{figure*}

We measure the line-of-sight velocity dispersion for the final
objects to compare them with the typical value of about
$10$~km\,s$^{-1}$ observed in classical dSph galaxies \citep{mun05}.

To measure the line-of-sight velocity dispersion of our models we
consider three different strategies.  First, $\sigma_{\rm 0,mean}$,
where we consider the velocity dispersion of all particles within a
projected radius of $10$~pc from the centre of the object.  The mean
value is calculated by considering the mean of the values obtained
along all three coordinate axis because the orientation of our objects
is unknown.  Further we determine the radial profile of the
line-of-sight velocity dispersion and fit a Plummer law to the data.
From this fit we determine the central value $\sigma_{\rm 0,fit}$.
Finally, we calculate the velocity dispersion of all particles within
a projected radius of $500$~pc -- $\sigma_{500}$.  The values of these
quantities can be found in Tab.~\ref{tab:vel1}.  In this section we
will focus on discussing $\sigma_{500}$ and its relation to the
initial parameters. 

In the top row of Fig.~\ref{fig:sigma03bin} we show the dependency of
$\sigma_{500}$ on the initial parameters of our models: $M_{500}$,
$R_{\rm h}$ and $R_{\rm sc}$.  In the top left panel we show the 
dependency on $M_{500}$.  As expected, velocity dispersion increases
with the mass of the DM halo.  As mass (potential energy) is related
to the square of the velocity (kinetic energy) we see this trend in
our models as well.  A bit puzzling are the results shown in the
middle and right panels.  Theory of virial equilibrium predicts that
\begin{eqnarray}
  \label{eq:vireq}
  \sigma^{2} & \sim & \frac{M_{\rm dyn}}{r_{\rm scale}},
\end{eqnarray}
i.e.\ the velocity dispersion should decrease with increasing
scale-length.  This relation is true using the total dynamical mass
(i.e.\ here the mass of the halo) and the velocity dispersion and
scale-length of the tracer population, i.e.\ in our case
$\sigma_{500}$ and $R_{\rm eff}$ of the final object.  As we have
shown in Fig.~\ref{fig:bright} that $R_{\rm eff}$ increases slightly
with $R_{\rm h}$ and strongly with $R_{\rm sc}$, we would expect a
decreasing trend in $\sigma_{500}$ with those two parameters as well.
Instead we see rather constant values for $\sigma_{500}$ as function
of $R_{\rm sc}$ and even a slight increase (but within the errors) as
function of $R_{\rm h}$. 

This behaviour can be explained by an increase of velocity streams at
larger scale-lengths, stemming from the dissolved star clusters.  We
will discuss these effects in a separate section below.  Finally, we
see no difference between cusped and cored haloes. 

For DM haloes with masses of $M_{500} = 10^{7}$~M$_{\odot}$, the
velocity dispersion of our final objects have similar values like the
ones measured for some classical dSph galaxies \citep{mun05}.  For
example: Carina, Leo II and Sextans have velocity dispersions around
$7.0$~km\,s$^{-1}$.  This is in fact an interesting result.  As
\citet{wal07} claimed that almost all dwarfs galaxies should have a
minimum DM halo mass of about $10^{7}$~M$_{\odot}$ within a
radius of $300$~pc, our models show that we can reach the same
velocity dispersions with a lower halo mass.  

Another consequence of our scenario is, that objects formed in
identical haloes can show slightly different velocity dispersions.
Otherwise we would have no error-bars attached to our data-points.
The reason for this will become clear with the discussion of the
results in the following subsection.  

\begin{figure*}
  \centering
  \epsfxsize=16cm
  \epsfysize=16cm
  \epsffile{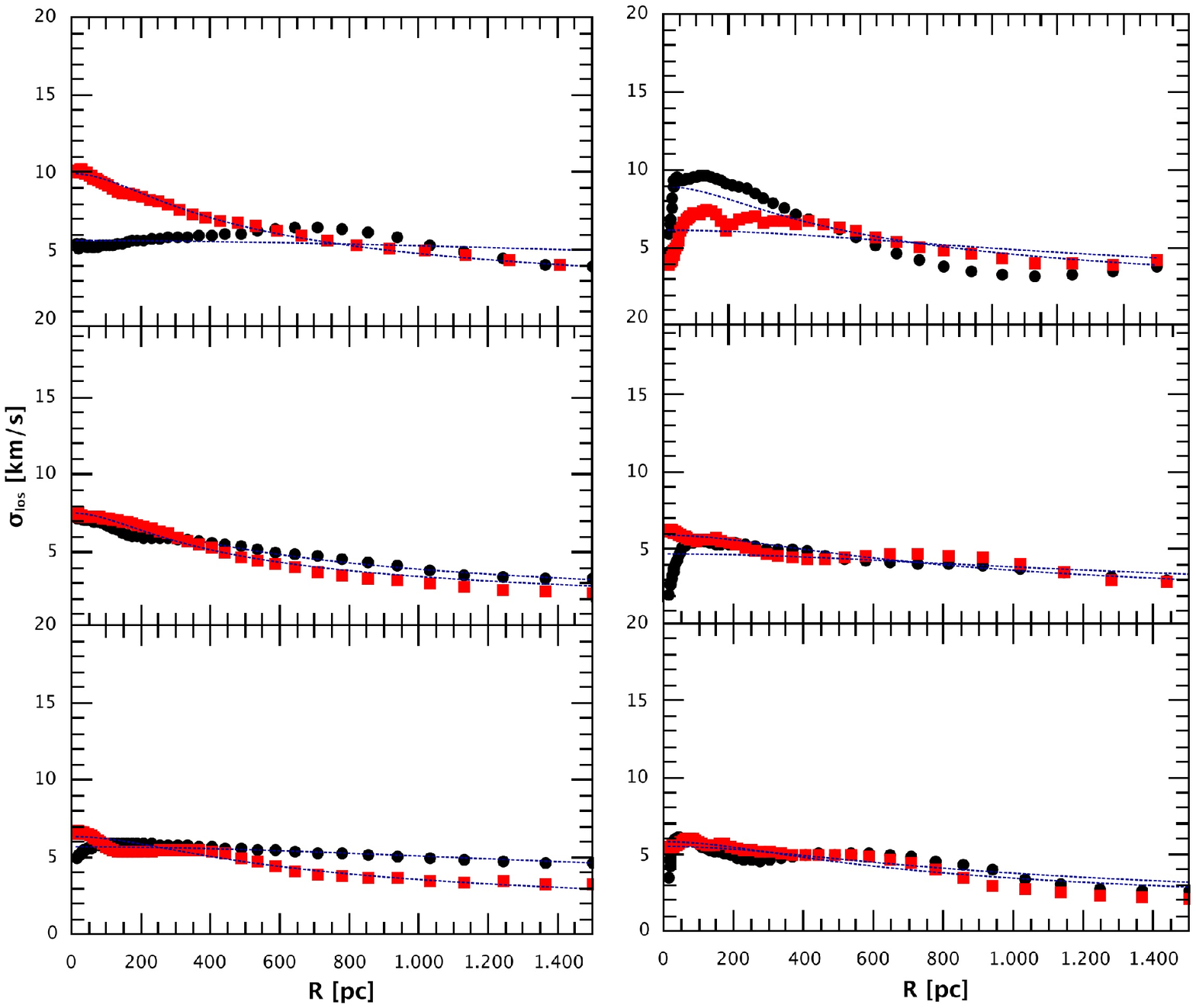}
  \caption{Line-of-sight velocity dispersion profiles for cusped (left
    panels) and cored (right panels) DM haloes for simulations with
    $M_{500} = 10^{7}$~M$_{\odot}$.  From top to bottom, the
    scale-lengths of the DM halo are $1.0$, $0.5$ and $0.25$~kpc,
    respectively.  Black and red colours correspond to simulations
    which initially have 15 or 30 star clusters.  Dashed lines are the
    fitting curves given by a Plummer fit to the data.  For both types
    of DM haloes the velocity dispersion profiles obtained are always
    more or less flat, out to large radii, as seen with dSph
    galaxies.}  
  \label{fig:vlos}
\end{figure*}

In Fig.\ref{fig:vlos}, we show the line-of-sight velocity dispersion
profiles for cusped (left panels) and cored (right panels) DM haloes
for some of our simulations whose $M_{500} = 10^{7}$~M$_{\odot}$.
These profiles have been obtained considering the mean value of the
line-of-sight velocity dispersion within concentric rings of varying
radius.  From top to bottom, the scale-lengths of the DM haloes are
$1.0$, $0.5$ and $0.25$~kpc, respectively.  For both types of DM
haloes the velocity dispersion profiles obtained are always more or
less flat.  

Looking at the outer part of the profiles we observe different types
of behavior in the dispersions.  We get outer profiles (beyond
$1$~kpc) where the velocity dispersion falls slowly in some cases,
while in others the dispersion stays flat.  We see a similar behavior
in the MW\'s dwarf galaxies.  In Sextans we see a slight drop in
velocity dispersion around 1 kpc, while Sculptor, Draco and Fornax
show flat profiles \citep[see fig.~2 in][]{wal07}.   
 
In the intermediate part (0.1 to 1 kpc) some of our simulations show
wiggles and bumps in the dispersion profile.  While the observers
always try to fit smooth curves, thus implicitly assuming that any
bumps seen in the profiles are merely due to statistical noise,
(e.g. Sculptor, Draco or Carina) in our formation scenario these
wiggles are real features.  These deviations are due to artifacts in 
velocity space, we dubbed fossil remnants (see the following section).
If our formation theory is true, the wiggles in the observed profiles
might not be because of errors but they might be real features. 

In the central part we see the same behavior as shown in the
multitude of MW's dwarf galaxies.  Some of our models have colder
cores, i.e.\ their dispersion profile has a central dip like in Sextans
or Draco \citep{mun05, wal07}.  Other models show a rising
central velocity dispersion.  We do not see a similar behaviour with
the classical dwarfs spheroidals of the MW.

These results show clearly that our models are well suited to
reproduce dSph galaxies.  We reproduce the dynamics of the different
dwarfs galaxies with our different models.  In our models, the bumps
in the observed profiles are not simply due to noise in the
observed data sets.  According to our formation theory they are a
natural by-product of the formation scenario proposed in this work.  
  
\subsubsection{Fossil remnants}
\label{sec:fr}

\begin{figure*}
  \centering
  \epsfxsize=16cm
  \epsfysize=12cm
  \epsffile{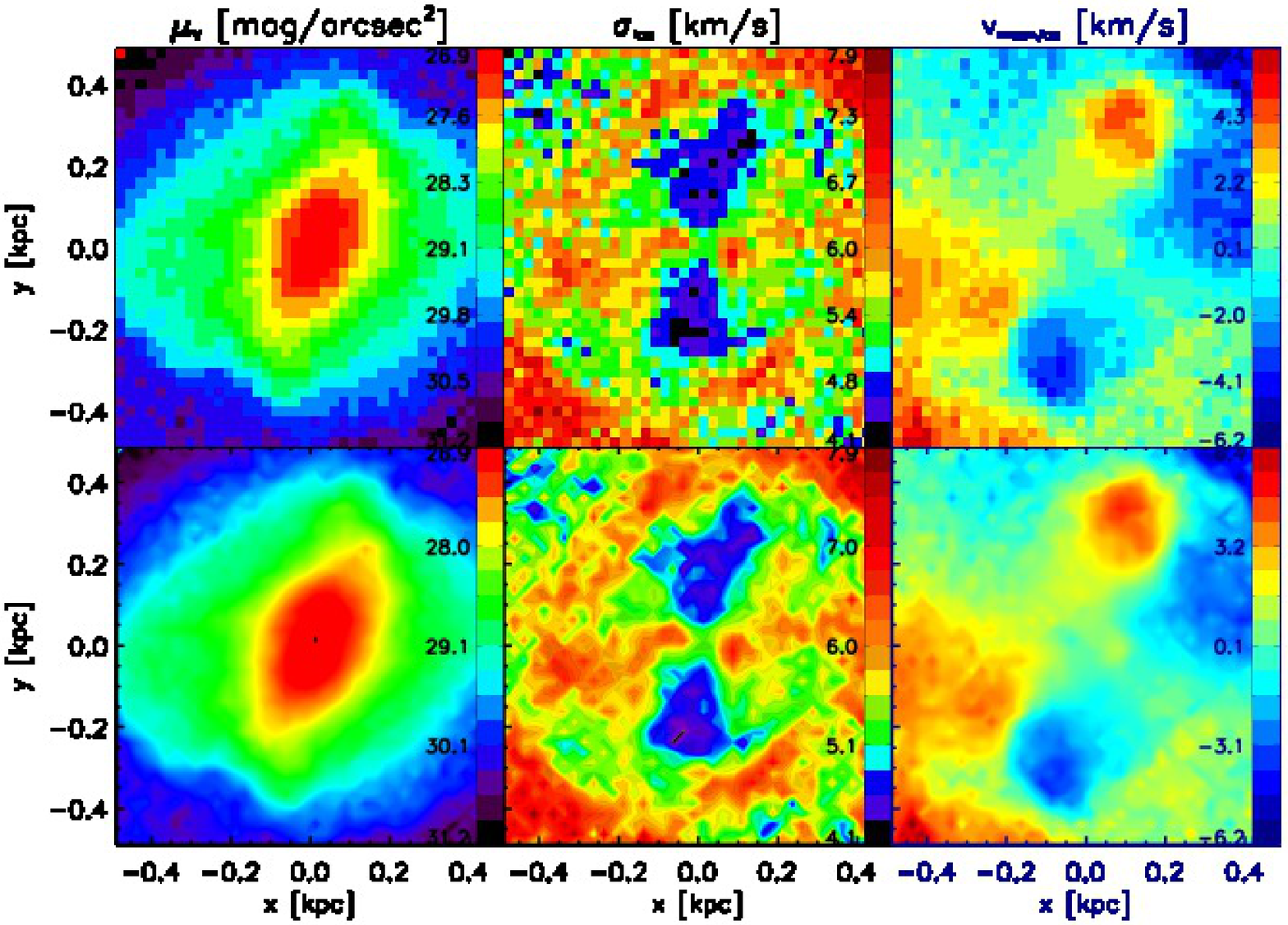}
  \caption{Shape and dynamics of simulation N015M107RH100RS025S30
    within $500$~pc as 2D pixel maps (top row) or as colour contours
    (bottom row).  The pixel size is $25$~pc.  We show the surface
    brightness of our object in the left panels (using an arbitrary
    mass-to-light ratio of $M/L = 1$ to convert masses in
    luminosities), the line of sight velocity dispersion calculated
    for each pixel in the middle panels and the mean velocity in each
    pixel in the right panels.}
  \label{fig:overn}
\end{figure*}

\begin{figure*}
  \centering
  \epsfxsize=16cm
  \epsfysize=08cm
  \epsffile{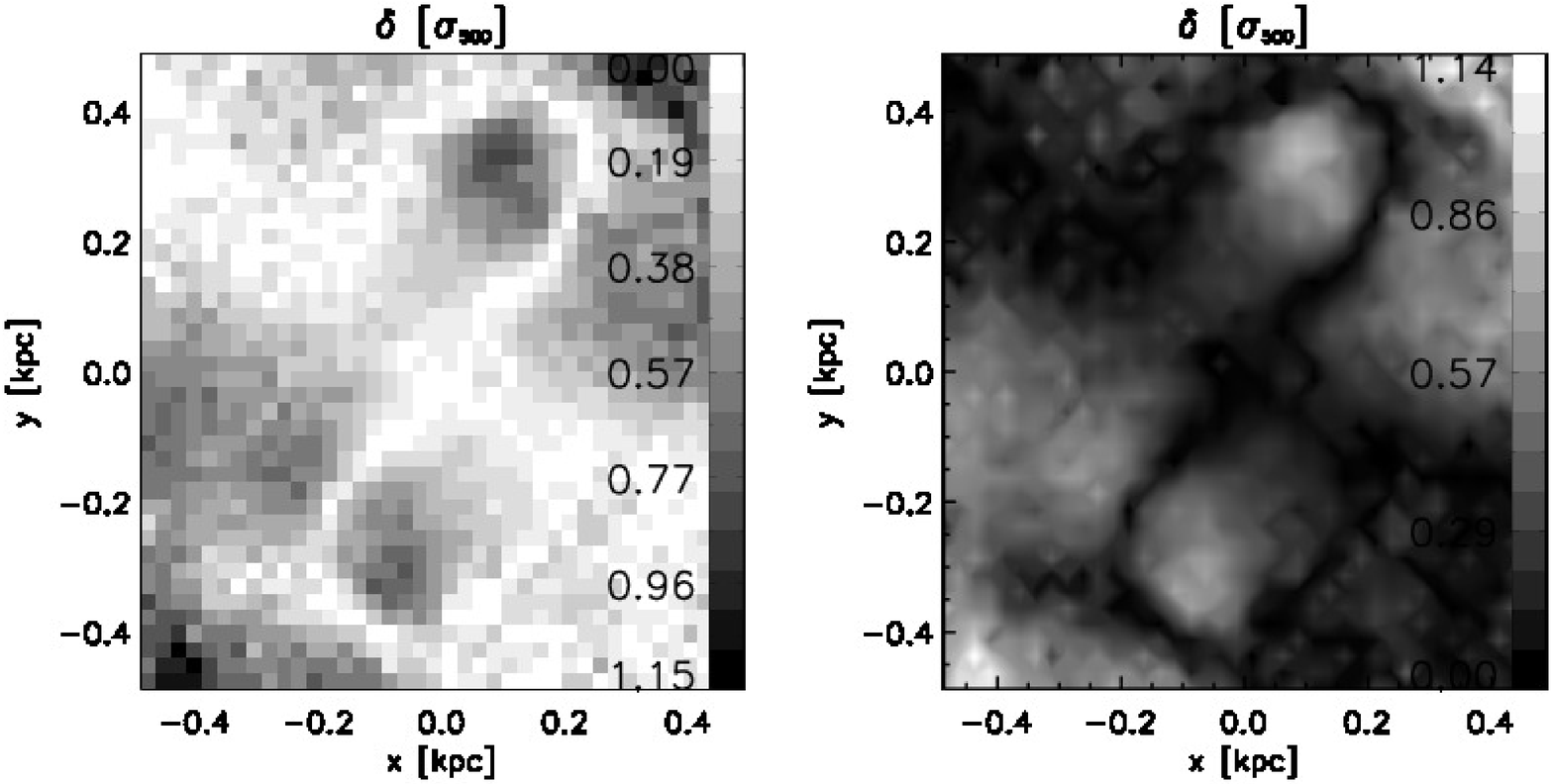}
  \caption{$\delta$-parameter distribution for the simulation
    N015M107RH100RS025S30.  Left panel shows the pixel map and right
    panel the contours now with inverted grey-scale.  The resolution
    is $25$~pc per pixel.}
  \label{fig:devn}
\end{figure*}

In Fig.~\ref{fig:overn}, we show the shape and dynamics of a
representative example of our simulations within $500$~pc as 2D pixel
maps.  As we are using more particles than actual stars we are able to
produce high resolution 2D pixel maps not only of the surface
brightness but also (as we know exactly the velocities of all
particles) of the internal dynamics of the system. 

In the left panels we show the surface brightness of the simulation 
N015M107RH100RS025S30 with a resolution of $25$~pc.  These panels
show, clearly, a centrally peaked and smooth density distribution for
the luminous component of our final object.  This distribution differs
from our fiducial model \citep{ass12}, indicating that the number of
initial star clusters, can affect the final properties of the dSph
galaxies.  However, the fact that a simulation with a lower number of
star clusters forms a smoother surface density distribution is not due
to some unknown and implausible physical effect, it is simply the
result of the strong randomness of our results.  The final details of
our models do not only depend on the initial parameters, they are also
strongly dependent on the random placements and orbits of the initial
star clusters.  As we are using only 15 to 60 star clusters in our
models, it is clear that the detailed results are highly dependent on
the random realisation of those few clusters.  In the previous
sections we tried to overcome this randomness by calculating mean
values taken from many simulations of the same type.

In the middle-column of Fig.~\ref{fig:overn}, we show the
line-of-sight velocity dispersion calculated for each pixel
separately.  We see regions with cold velocities of around
$4.5$~km\,s$^{-1}$ in two opposing regions, almost but not perfectly
aligned with the major axis of the dwarf.  In the outskirts the
velocity dispersion increases.  But the distribution of the velocity
dispersion is far from being smooth. One can, clearly, observe
substructures that are not present in the surface brightness plots.
While the DM halo was able to erase the substructures in position
space, substructures survive in velocity space.  We refer to these
substructures in velocity space, that remain after the formation
process of our objects, as 'fossil remnants'.  In all of our
simulations we observe these substructures in velocity space.

In the right-column panels of Fig.~\ref{fig:overn}, we show the mean
velocity of all particles within a pixel.  These figures show regions
of coherent streaming motions with velocity differences of up to
$4$~km\,s$^{-1}$. The presence of streaming motions means that there
are groups of stars, of dissolved clusters, that are trapped in the
potential well of the halo at uniform motion.  It is possible to
observe at least two ring-like structures of opposing radial
velocities. This type of scenario has been considered, recently, to
explain the velocity distribution of the UMi dSph galaxy
\citep{kle03}.  The reason why we should expect to observe these ring
structures, in velocity space, is that the dSph galaxies are faint
objects with a low number density of stars, therefore even a few stars
on a similar orbit can produce a detectable signal.  Of further
interest is that the features in the mean velocity do not necessarily
coincide with features we see in the dispersion map.  In
\citet{ass12}, we showed that the boosted velocity dispersions stem
from distributions which are highly non-Gaussian, because they are in
fact the overlay of stars on different orbits of the initial star
clusters. 

Our model shows that ring structures are present even after 10 Gyr,
and this is physically expected due to the long relaxation times of
the stars in dwarf galaxies.  That ring structures are formed from
star clusters on more circular orbits, that spread their stars slowly
along the orbit.  In the case that the star cluster have radial orbits
they will tend to disperse, since they will interact with other
clusters in the centre of the halo or with the central cusp.  Thus,
the ring structure should not be within the central area of the galaxy
and should expand during the evolution of the dwarf galaxy. 

An important consequence of having streaming motions in dwarf
galaxies is the fact that it implies, if not properly taken into
account, an overestimation of the velocity dispersion.  This is the
reason why we need less DM (than models which are feature-less and in
equilibrium) in our scenario to explain the velocity dispersions
typical of observed dwarf spheroidal galaxies.

To quantify the strength of the streaming motions in the objects, we
introduce the $\delta$-parameter:
\begin{eqnarray}
  \label{eq:delta}
  \delta & = & \frac{\left| v_{\rm mean, pixel} - v_{\rm mean,
      object} \right|}{\sigma_{500}}.
\end{eqnarray}
The advantage is that it shows regions where the deviation of the mean
velocity is distinct from zero.  Higher values of $\delta$ correspond 
to regions where the streaming motions are strong.  If no streaming
motions are present, the $\delta$-parameter is randomly distributed
without correlations.  Also it is easily shown by means of basic
statistics or Monte Carlo simulations \citep{fel12} that the
maximum value of $\delta$ due to random effects only depends on the
number of velocities you use to calculate the mean.  Random deviations
of the calculated mean velocity from the 'true value' due to low
number statistics on a three-sigma level lead to $\delta$  values of
$1.25$ for $10$~stars, $0.4$ for $100$~stars and $0.14$ for
$1000$~stars, i.e.\ measurements.  As said before we have an
over-sampling of particles in our simulations and the pixels in
Figs.~\ref{fig:devn} and \ref{fig:overn} contain 300 particles or
more.  So all the velocity features we see are real and not due to low
numbers.  In these high resolution plots we have 300 to 3000 particles
in each pixel so as a conservative number whenever $\delta > 0.3$, the
streaming motions are real and not due to random sampling.

In Fig.~\ref{fig:devn} we show the $\delta$ distribution for the same
simulation as shown in Fig.~\ref{fig:overn}.  Once again a clear sign
that we are dealing with coherent motions and not random deviations is
that we see regions spanning many pixel showing similar high $\delta$
values.  The imaginary figure eight with low values has no special
meaning and is rather a matter of contingency.

The maximum value of $\delta$ found in a single pixel within a region
of radius $500$~pc we call $\delta_{500}$.  It is a measure of the
strength of the velocity deviations in our models and allows us to
compare the simulations with each other.  The values are given in
Tab.~\ref{tab:vel1}. 

If we analyse the dependencies of $\delta_{500}$ we see in the bottom
panels of Fig.~\ref{fig:sigma03bin} that we have no dependency of this
parameter with the mass of the halo (except maybe that we see less
scatter in the results at higher masses) or with the scale-length of
the SC distribution.  We see that the majority of symbols representing
cored Plummer haloes are above the corresponding NFW haloes meaning
that cusped halo profiles seem to be better in erasing kinematical
substructure (i.e.\ fossil remnants).  But that would imply that we
should also see a dependency with the halo scale-length, meaning
larger scale-lengths lead to higher values.  This behaviour is not
visible in our data, although three out of four data-points for
$R_{\rm h} = 1.0$~kpc lie above $\delta_{500} = 1.0$, while for the
other scale-lengths only one out of four is above this value.  

We also can not identify a dependency on the number of SCs as long as
the numbers are low (i.e.\ $N_{0} = 15, 30$).  If we use $N_{0} = 60$,
we see the lowest value of $\delta_{500}$.  We conclude that we would
need to distribute the luminous mass into a real high number of
low-mass SCs to really see a strong dependency on $N_{0}$, meaning
that $\delta_{500}$ tends to zero, i.e.\ is below the stochastic
noise. 

While we have shown in \citet{ass12} that in reality even with more
than 2000 radial velocities observed it is still almost impossible to
detect the fossil remnants in low-resolution 2D plots, they have
visible counterparts in the 1D bumps and wiggles of the radial
profiles.  Of course we average the high discrepancies out by taking
means over a whole radial ring but some small deviations remain
visible.  As already pointed out in \citet{ass12} these bumps and
wiggles are present in the observational data but are regarded as
statistical noise as they are smaller than the errors.  We see the
same behaviour in our models and these deviations are real.


\section{Discussion}
\label{sec:dis}

 Our initial conditions assume that the star clusters are initially distributed according to a non-rotating Plummer profile within the dSph halo. Whether this is a realistic assumption or whether we should instead consider an initial distribution with non-zero angular momentum  is an open question. However, we note that the hierarchical merger process through which dSphs acquire their dark matter and gas at high redshift are very violent and it is therefore likely that star formation, when it occurs, takes place before the gas distribution has time to settle into a rotationally supported disc. Even if such a disc were to form, observations show that the molecular and atomic gas of disk galaxies generally has velocity dispersions higher than $5$ ~kms$^{-1}$ \citep{sta89,tam09} which is taken to indicate the presence of gas turbulence. Given that such turbulent gas velocities are similar to the velocity dispersion  in a dSph, any disc which formed would likely have a scale-height to scale-radius of order unity and therefore not be very dissimilar to our assumed spherical initial conditions~\citep[e.g.][]{rpv06}

We also note that the Leo~T dwarf galaxy has a gas disc which displays negligible rotation and therefore constitutes an example of a system with little or no primordial angular momentum, but sufficient gas to support further star formation. Our models are thus directly relevant for an object such as Leo~T, as well as any systems with more spheroidal gas distributions. Given that the star clusters in our simulations are initially expanding due to gas expulsion, it is possible that our models would yield dSphs with little or no angular momentum, even if the initial configuration of the clusters were disc-like and contained some angular momentum.  We intend to explore such initial conditions in a future paper.

Finally, we note that given the relatively short crossing time within the dSph halo, as well as the low star formation efficiency within the star clusters, the final distribution of the luminous component of our remnants is relatively insensitive to the initial spatial distribution we assume. In particular, Plummer profiles do not yield good fits to the final stellar distributions of our final objects, demonstrating that the merger process has erased much of the original configuration. Thus our assumption of a particular form for the distribution of our clusters has little impact on the properties of the dSphs we form.

The question remaining is: are such velocity structures, as predicted
by our formation scenario, present and observable in real dSph
galaxies.  If we have sufficient radial velocity measurements we
can calculate similar 2D maps like in our simulations.  We do
not expect, in the near future, to have a sufficient pixel-resolution
with real observations to see coherent structures, i.e.\ many pixels
with the same motion next to each other.  But any sign of symmetry of
pixels with high $\delta$ would be a good indicator.  If we look at
the maximum $\delta$-values obtained in our simulations, we deduce,
that as soon as we have observational pixel maps with 50-100 stars per
pixel almost all the 'real' $\delta$-values from the coherent motions
in our simulations should be detectable.  The strongest values of
$\delta$, we detect in our simulations, could even be visible if
small enough pixels with 10-20 stars only are observed.

Since we are considering that the final objects of our models resemble
dSph galaxies, and thus that our formation scenario corresponds to the
formation scenario of dSph galaxies, it is legitimate to ask if such
streaming motions could be present in known dSph galaxies.  Some of
the classical dSph galaxies of the Local Group have now more than 2000
radial velocities measured and the first low resolution pixel maps of
the velocities should be feasible to produce in the near future.

It is known that Ursa Major II and Hercules show strong velocity
gradients of 7 to 14~km\,s$^{-1}$.  If these measurements are not due
to tidal distortion, we estimate that these gradients correspond to
values of $\delta > 0.5$, which is in accordance with most of the
$\delta$-values obtained in our models.

Another important point to observe from our results is the fact that
low SFE (SFE $\leq$ 30\%) is necessary to obtain our objects.  That
is, low SFE lead to subsequent dissolution of the initial star
clusters.  Only in this case we get sufficiently low phase space
densities of the luminous components that resemble dSph galaxies.
Merging star clusters without expansion always lead to compact
objects, even without the presence of a DM halo \citep{fel02}.  

As mentioned before, there are simulations where some of the initial
star clusters survive.  In these cases, the star clusters escaped from
the dissolution process by experiencing a compressing force in the
central region of the halo, allowing the cluster to stay bound.  We
find these star clusters on orbits which never leave the central
region of the luminous component of the dwarf galaxies.

The survival of the star clusters seems to be completely random in our
simulations, that is, it does not seem to depend on the initial
parameters $M_{500}$, $R_{\rm h}$ or $R_{\rm sc}$. But, we do see more
survivors in cored than in cusped haloes.  A flat central density
structure seems to act in favour of constructive compression than of
destructive tidal forces. 

Still this mechanism can not explain the presence of globular clusters
associated with dSph galaxies like Fornax.  We strongly believe that
those clusters, outside or in the outer parts of the luminous
components, must have formed with high SFEs, which allowed them to
survive in the first place.  We see this behaviour in our simulations
with an artificial high SFE of $60$~per cent, in which all of our SCs
are still present even after $10$~Gyr of evolution, i.e.\ the tidal
forces of the dSph galaxy are not sufficient to disrupt those clusters
completely (Note, that our code can not model any internal evolution
due to two-body effects inside the SCs).  I.e.\ our simulations
provide a straightforward solution to why some dSph galaxies have
orbiting star clusters and why some (most) of them do not.

Finally, we have to admit that most of our trends are still based on
low number statistics.  Even though this study is based on more than 60
simulations, using random placements of only 15 or 30 star clusters
can largely affect the outcome of a single simulation.  We would need
a much larger sample sizes of simulations for each set of initial
parameters to be more confident in our trends.  Nevertheless, those
trends would still only provide a prediction in a statistical sense
and we would be able to deduce the most likely initial parameters of a
real dSph galaxy only.  


\section{Conclusions}
\label{sec:conc}

In this work we propose and test a new scenario for the formation of
dSph galaxies.  The scenario is based on two standard theories.  We
consider that stars never form in isolation but in star clusters which
suffer from gas expulsion.  We also base our model on the widely accept
$\Lambda$-CDM cosmological model, where it is assumed that small DM
haloes form first.  We propose that the dynamical evolution of
star clusters, i.e.\ their dissolution due to gas expulsion in the
centre of small DM haloes, may explain the formation of classical dSph
galaxies.  In our scenario we simulate a DM halo with star clusters
inside.  These star clusters have low SFE and, thus, are designed to
dissolve inside the DM halo forming the luminous components of the
dSph galaxy.  We follow the evolution of the star clusters within the
DM halo for $10$ Gyr and measure the properties of the objects
formed.

In this work we perform an investigation of the vast possible
parameter space and study how the morphology, luminosity and velocity
space of the final objects are affected by the initial conditions of
the simulations. In our experiments we take cusped and cored DM haloes
into account, and study, for each type of DM halo, how different mass
distributions, number of interacting star clusters, SFEs
star formation efficiencies and initial distributions of the
interacting star clusters can affect the formation process of the dSph
galaxies.

Regarding the light distribution we see the following trends in our
simulations: a decrease of effective surface brightness with larger
halo scale-lengths and an increase of the effective radius of the
luminous component with increasing scale-length of the initial SC-
and/or gas-distribution.  We can identify two rule by thumbs.  To
reach similar low brightnesses as seen in Carina, Draco and Scupltor
our models favour larger scale-lengths of the halo, i.e.\ $> 500$~pc.
Another trend is that the effective radius of the luminous object is
almost twice as large as the initial scale-length of the SC
distribution. 

Even though our model starts out initially very 'clumpy', i.e.\ the
stars are distributed in star clusters, we obtain smooth models at the
end of the evolution.  Even though relaxation times are long, the
stars settle into a more or less smooth object.  Of course we find
that cored haloes are better in preserving some 'clumpiness' than
cusped haloes.  

The objects have internal structures within $500$~pc that
have an observable luminosity comparable to the luminosity of dSph
galaxies.  They also have similar morphological parameters.  
One interesting result from our simulations is that even 10~Gyr of
evolution is not enough to destroy all substructures that stem from
the formation process of our models.   

Even though we start out with spherical distributions of the star
clusters our final luminous object has ellipticities similar to those
observed with the classical dwarf galaxies.  Again, we see a trend
that larger scale-lengths of the haloes lead to higher ellipticities. 

In velocity space, the final objects have overall properties similar to
typical classical dSph galaxies.  We show that the velocity dispersions
are around $5$--$10$~km\,s$^{-1}$ for halo masses of $M_{500} =
10^{7}$~M$_{\odot}$ and that the velocity dispersion profiles do not
fall sharply at large radii, which is an indication of DM dominated
objects as it is the case of dSph galaxies.

As a final result, we observe in the velocity space of our models some
fossil remnants of the formation process, that give predictions for 
dynamical observations that could be used for validating our scenario
about the formation of the dSph galaxies.  We show the existence of
streaming motions even after 10~Gyr of evolution, irrespective of halo
mass, shape or scale-length.  With the larger-than-life resolution we
have in our simulations, we can visualise these streams in form of
2D-contour plots of the dispersion and the mean velocity.
Observationally, this is still impossible.  But, we show that these
velocity deviations are still visible, if averaged in radial bins.
Then, these deviations produce wiggles and bumps in the profile,
similar to the observed profiles of the classical dwarfs.
\\

{\bf Acknowledgments:}
PA thanks the support of CONICYT PhD scholarship, BASAL PFB-06/2007,
PROYECTO FONDAP 15010003 and the travel grant MECESUP-FSM0605.  We
also thank W. Dehnen for his help with the NFW profiles and P. Kroupa
for enlightening discussions.  MF acknowledges financial support of
FONDECYT grant nos.\ 1095092 and 1130521 and BASAL PFB-06/2007.  MIW
acknowledges the Royal Society for support.  RS acknowledges financial
support of FONDECYT grant no.\ 3120135.

\appendix
\onecolumn

\section{Tables of all results}
\label{sec:tables}

\begin{center}

\begin{longtable}{lcccccccccc} 
  \caption{Table summarizing the initial parameters of our
    simulations.  In the first column we give a running number which
    we use in the following tables.  In the second column we give the
    name of each simulation according to the rule introduced in
    Table~\ref{tab:conf}.  The third column shows what type of DM
    profile is used; P stands for a Plummer profile and N for a NFW
    halo. The fourth and fifth columns show the mass of the DM halo
    within $500$~pc ($M_{500}$) and the total mass $M_{\rm tot}$ out
    to the virial radius, respectively.  In case of a Plummer profile
    the latter is the cut-off radius.  The sixth column shows the
    scale-length of the halo $R_{\rm h}$; in case of a Plummer profile
    this is the Plummer radius.  The seventh column gives the virial
    radii for the NFW haloes or the cut-off radii of the Plummer
    distribution.  In case of the Plummer profile the halo is
    truncated at five Plummer radii.  The concentration parameter $c =
    R_{\rm vir} / R_{\rm h}$ is only given for NFW haloes.  The last
    three columns show the number of initial star clusters $N_{0}$,
    the scale-length (i.e.\ Plummer radius) of the distribution of
    star clusters $R_{\rm sc}$ and finally the star formation
    efficiency SFE used for each of the star clusters.} \\
  \hline
  \label{tab:init-cond}
  Number & Simulation & P/N & $M_{500}$ & $M_{\rm tot}$ & $R_{\rm h}$ &
  $R_{vir}$ & c & $N_{0}$ & $R_{\rm sc}$ & SFE \\
  & & &[M$_{\odot}$] & [M$_{\odot}$] & [kpc] & [kpc] & & & [kpc] & \\
  \hline
  \endfirsthead
  \hline    
  Number & Simulation & P/N & $M_{500}$ & $M_{\rm tot}$ & $R_{\rm h}$ &
  $R_{vir}$ & c & $N_{0}$ & $R_{\rm sc}$ & SFE \\
  & & &[M$_{\odot}$] & [M$_{\odot}$] & [kpc] & [kpc] & & & [kpc] & \\
  \hline
  \endhead
  \hline
  \endfoot
  $01$ & N015M107RH025RS025S30 & N & $1 \times 10^{7}$ & 
  $5.8 \times10^{7}$ & $0.25$ & $8.0$ & $32$ & $15$ & $0.25$ & $0.30$ \\ 
  $02$ & P015M107RH025RS025S30 & P & $1 \times 10^{7}$ & 
  $1.4\times10^{7}$ & $0.25$ & $1.25$ & -- & $15$ & $0.25$ & $0.30$ \\
  $03$ & N030M107RH025RS025S30 & N & $1 \times 10^{7}$ & 
  $5.8 \times10^{7}$ & $0.25$ & $8.0$ & $32$ & $30$ & $0.25$ & $0.30$ \\
  $04$ & P030M107RH025RS025S30 & P & $1 \times 10^{7}$ & 
  $1.4\times10^{7}$ & $0.25$ & $1.25$ & -- & $30$ & $0.25$ & $0.30$ \\
  $05$ & N015M107RH050RS025S30 & N & $1 \times 10^{7}$ & 
  $1.1 \times10^{8}$ & $0.5$ & $9.8$ & $20$ & $15$ & $0.25$ & $0.30$ \\
  $06$ & P015M107RH050RS025S30 & P & $1 \times 10^{7}$ & 
  $2.8\times10^{7}$ & $0.5$  & $2.5$ & -- & $15$ & $0.25$ & $0.30$ \\
  $07$ & N030M107RH050RS025S30 & N & $1 \times 10^{7}$ & 
  $1.1 \times10^{8}$ & $0.5$ & $9.8$ & $20$ & $30$ & $0.25$ & $0.30$ \\
  $08$ & P030M107RH050RS025S30 & P & $1 \times 10^{7}$ & 
  $2.8\times10^{7}$ & $0.5$  & $2.5$ & -- & $30$ & $0.25$ & $0.30$ \\
  $09$ & N015M107RH100RS025S30 & N & $1 \times 10^{7}$ & 
  $1.3 \times10^{8}$ & $1.0$  & $13$ & $13$ & $15$ & $0.25$ & $0.30$ \\
  $10$ & P015M107RH100RS025S30 & P & $1 \times 10^{7}$ & 
  $1.1\times10^{8}$ & $1.0$  & $5.0$ & -- & $15$ & $0.25$ & $0.30$ \\
  $11$ & N015M107RH100RS025S15 & N & $1 \times 10^{7}$ & 
  $1.3 \times10^{8}$ & $1.0$ & $13$ & $13$ & $15$ & $0.25$ & $0.15$ \\
  $12$ & P015M107RH100RS025S15 & P & $1 \times 10^{7}$ & 
  $1.1\times10^{8}$ & $1.0$ & $5.0$ & -- & $15$ & $0.25$ & $0.15$ \\
  $13$ & N015M107RH100RS025S60 & N & $1 \times 10^{7}$ & 
  $1.3 \times10^{8}$ & $1.0$ & $13$ & $13$ & $15$ & $0.25$ & $0.60$ \\
  $14$ & P015M107RH100RS025S60 & P & $1 \times 10^{7}$ & 
  $1.1\times10^{8}$ & $1.0$ & $5.0$ & -- & $15$ & $0.25$ & $0.60$ \\
  $15$ & N015M107RH100RS050S30 & N & $1 \times 10^{7}$ & 
  $1.3 \times10^{8}$ & $1.0$ & $13$ & $13$ & $15$ & $0.5$ & $0.30$ \\
  $16$ & P015M107RH100RS050S30 & P & $1 \times 10^{7}$ & 
  $1.1\times10^{8}$ & $1.0$ & $5.0$ & -- & $15$ & $0.5$ & $0.30$ \\
  $17$ & N015M107RH100RS100S30 & N & $1 \times 10^{7}$ & 
  $1.3 \times10^{8}$ & $1.0$ & $13$ & $13$ & $15$ & $1.0$ & $0.30$ \\
  $18$ & P015M107RH100RS100S30 & P & $1 \times 10^{7}$ & 
  $1.1\times10^{8}$ & $1.0$ & $5.0$ & -- & $15$ & $1.0$ & $0.30$ \\
  $19$ & N030M107RH100RS025S30 & N & $1 \times 10^{7}$ & 
  $1.3 \times10^{8}$ & $1.0$  & $13$ & $13$ & $30$ & $0.25$ & $0.30$ \\
  $20$ & P030M107RH100RS025S30 & P & $1 \times 10^{7}$ & 
  $1.1\times10^{8}$ & $1.0$  & $5.0$ & -- & $30$ & $0.25$ & $0.30$ \\
  $21$ & N030M107RH100RS025S15 & N & $1 \times 10^{7}$ &
  $1.3 \times10^{8}$ & $1.0$ & $13$ & $13$ & $30$ & $0.25$ & $0.15$ \\
  $22$ & P030M107RH100RS025S15 & P & $1 \times 10^{7}$ & 
  $1.1\times10^{8}$ & $1.0$ & $5.0$ & -- & $30$ & $0.25$ & $0.15$ \\
  $23$ & N030M107RH100RS025S60 & N & $1 \times 10^{7}$ & 
  $1.3 \times10^{8}$ & $1.0$ & $13$ & $13$ & $30$ & $0.25$ & $0.60$ \\
  $24$ & P030M107RH100RS025S60 & P & $1 \times 10^{7}$ & 
  $1.1\times10^{8}$ & $1.0$ & $5.0$ & -- & $30$ & $0.25$ & $0.60$ \\
  $25$ & N060M107RH100RS025S30 & N & $1 \times 10^{7}$ & 
  $1.3 \times10^{8}$ & $1.0$ & $13$ & $13$ & $30$ & $0.25$ & $0.30$ \\
  $26$ & N030M107RH100RS050S30 & N & $1 \times 10^{7}$ & 
  $1.3 \times10^{8}$ & $1.0$ & $13$ & $13$ & $30$ & $0.5$ & $0.30$ \\
  $27$ & P030M107RH100RS050S30 & P & $1 \times 10^{7}$ & 
  $1.1\times10^{8}$ & $1.0$ & $5.0$ & -- & $30$ & $0.5$ & $0.30$ \\
  $28$ & N030M107RH100RS100S30 & N & $1 \times 10^{7}$ &
  $1.3 \times10^{8}$ & $1.0$ & $13$ & $13$ & $30$ & $1.0$ & $0.30$ \\
  $29$ & P030M107RH100RS100S30 & P & $1 \times 10^{7}$ & 
  $1.1\times10^{8}$ & $1.0$ & $5.0$ & --& $30$ & $1.0$ & $0.30$ \\
  $30$ & N015M407RH025RS025S30 & N & $4 \times 10^{7}$ & 
  $2.8\times10^{8}$ & $0.25$ & $14$ & $54$ & $15$ & $0.25$ & $0.30$ \\
  $31$ & P015M407RH025RS025S30 & P & $4 \times 10^{7}$ & 
  $5.6\times10^{7}$ & $0.25$ & $1.25$ & -- & $15$ & $0.25$ & $0.30$ \\
  $32$ & N030M407RH025RS025S30 & N & $4 \times 10^{7}$ & 
  $2.8\times10^{8}$ & $0.25$ & $14$ & $54$ & $30$ & $0.25$ & $0.30$ \\
  $33$ & P030M407RH025RS025S30 & P & $4 \times 10^{7}$ & 
  $5.6\times10^{7}$ & $0.25$ & $1.25$ & -- & $30$ & $0.25$ & $0.30$ \\
  $34$ & N015M407RH050RS025S30 & N & $4 \times 10^{7}$ & 
  $5.3\times10^{8}$ & $0.5$  & $17$ & $33$ & $15$ & $0.25$ & $0.30$ \\
  $35$ & P015M407RH050RS025S30 & P & $4 \times 10^{7}$ & 
  $1.1\times10^{8}$ & $0.5$  & $2.5$ & -- & $15$ & $0.25$ & $0.30$ \\
  $36$ & N030M407RH050RS025S30 & N & $4 \times 10^{7}$ & 
  $5.3\times10^{8}$ & $0.5$  & $17$ & $33$ & $30$ & $0.25$ & $0.30$ \\
  $37$ & P030M407RH050RS025S30 & P & $4 \times 10^{7}$ & 
  $1.1\times10^{8}$ & $0.5$  & $2.5$ & -- & $30$ & $0.25$ & $0.30$ \\
  $38$ & N015M407RH100RS025S30 & N & $4 \times 10^{7}$ & 
  $1.2\times10^{9}$ & $1.0$  & $22$ & $22$ & $15$ & $0.25$ & $0.30$ \\
  $39$ & P015M407RH100RS025S30 & P & $4 \times 10^{7}$ & 
  $4.5\times10^{8}$ & $1.0$  & $5.0$ & -- & $15$ & $0.25$ & $0.30$\\
  $40$ & N030M407RH100RS025S30 & N & $4 \times 10^{7}$ & 
  $1.2\times10^{9}$ & $1.0$  & $22$ & $22$ & $30$ & $0.25$ & $0.30$ \\
  $41$ & P030M407RH100RS025S30 & P & $4 \times 10^{7}$ & 
  $4.5\times10^{8}$ & $1.0$  & $5.0$ & -- & $30$ & $0.25$ & $0.30$ \\
  $42$ & N015M108RH025RS025S30 & N & $1 \times 10^{8}$ & 
  $7.8\times10^{8}$ & $0.25$ & $19$ & $76$ & $15$ & $0.25$ & $0.30$ \\
  $43$ & P015M108RH025RS025S30 & P & $1 \times 10^{8}$ & 
  $1.4\times10^{8}$ & $0.25$ & $1.25$ & -- & $15$ & $0.25$ & $0.30$ \\
  $44$ & N030M108RH025RS025S30 & N & $1 \times 10^{8}$ & 
  $7.8\times10^{8}$ & $0.25$ & $19$ & $76$ & $30$ & $0.25$ & $0.30$ \\
  $45$ & P030M108RH025RS025S30 & P & $1 \times 10^{8}$ & 
  $1.4\times10^{8}$ & $0.25$ & $1.25$ & -- & $30$ & $0.25$ & $0.30$ \\
  $46$ & N015M108RH050RS025S30 & N & $1 \times 10^{8}$ & 
  $1.5\times10^{9}$ & $0.5$  & $24$ & $47$ & $15$ & $0.25$ & $0.30$ \\
  $47$ & P015M108RH050RS025S30 & P & $1 \times 10^{8}$ & 
  $2.8\times10^{8}$ & $0.5$  & $2.5$ & -- & $15$ & $0.25$ & $0.30$ \\
  $48$ & N030M108RH050RS025S30 & N & $1 \times 10^{8}$ & 
  $1.5\times10^{9}$ & $0.5$  & $24$ & $47$ & $30$ & $0.25$ & $0.30$ \\
  $49$ & P030M108RH050RS025S30 & P & $1 \times 10^{8}$ & 
  $2.8\times10^{8}$ & $0.5$  & $2.5$ & -- & $30$ & $0.25$ & $0.30$ \\
  $50$ & N015M108RH100RS025S30 & N & $1 \times 10^{8}$ & 
  $3.5\times10^{9}$ & $1.0$  & $31$ & $31$ & $15$ & $0.25$ & $0.30$ \\
  $51$ & P015M108RH100RS025S30 & P & $1 \times 10^{8}$ & 
  $1.1\times10^{9}$ & $1.0$  & $5.0$ & -- & $15$ & $0.25$ & $0.30$ \\
  $52$ & N030M108RH100RS025S30 & N & $1 \times 10^{8}$ & 
  $3.5\times10^{9}$ & $1.0$  & $31$ & $31$ & $30$ & $0.25$ & $0.30$ \\
  $53$ & P030M108RH100RS025S30 & P & $1 \times 10^{8}$ & 
  $1.1\times10^{9}$ & $1.0$  & $5.0$ & -- & $30$ & $0.25$ & $0.30$ \\
\end{longtable}

\begin{longtable}{lccccccc}
  \caption{Results of the surface density profiles fitted to each
    simulation.  The first column gives the number of the simulation
    according to Tab.~\ref{tab:init-cond}.  The next two columns give
    the central density $\Sigma_{\rm 0,k}$ and the core radius $R_{\rm
      c}$ for a King profile fit without a tidal radius ($
    \Sigma(R)=\frac{\Sigma_{0}}{1+(\frac{R}{R_{\rm c}})^{2}} $).
    Column four and five show the fitting parameters of a Plummer
    profile, i.e.\ central density $\Sigma_{\rm 0,p}$ and Plummer
    radius $R_{\rm pl}$.  The last three columns show the parameters
    of a S\'ersic profile: the surface density at the effective radius
    $\Sigma_{\rm eff}$, the power-law index $n$ and the effective
    radius $R_{\rm eff}$.} \\
  \hline
  \label{tab:res1}
  & King & & Plummer & & S\'ersic & & \\
  No.\ & $\Sigma_{\rm 0,k}$ & $R_{\rm c}$ & $\Sigma_{\rm 0,p}$ &
  $ R_{\rm pl}$ & $\Sigma_{\rm eff}$ & $n$ & $R_{\rm eff}$ \\
  & [M$_{\odot}$pc$^{-2}$] & [pc] & [M$_{\odot}$pc$^{-2}$] & [pc]
  & [M$_{\odot}$pc$^{-2}$] &  & [pc] \\
  \hline
  \endfirsthead
  \hline
  & King & & Plummer & & S\'ersic & & \\
  No.\ & $\Sigma_{\rm 0,k}$ & $R_{\rm c}$ & $\Sigma_{\rm 0,p}$ &
  $ R_{\rm pl}$ & $\Sigma_{\rm eff}$ & $n$ & $R_{\rm eff}$ \\
  & [M$_{\odot}$pc$^{-2}$] & [pc] & [M$_{\odot}$pc$^{-2}$] & [pc]
  & [M$_{\odot}$pc$^{-2}$]  &  & [pc] \\
  \hline
  \endhead
  01 & $2.17 \pm 0.02$ & $110 \pm 2$ & -- & -- &
  $7.2 \pm 0.1$ & $1.13 \pm 0.04$ & $280 \pm 10$ \\
  02 & $2.08 \pm 0.04$ & $190 \pm 10$ & $2.0 \pm 0.1$ & $300 \pm 10$ & 
  $5.5 \pm 0.1$ & $0.45 \pm 0.03$ & $187 \pm 7$\\
  03 & $1.60 \pm 0.02$ & $180 \pm 4$ & $1.56 \pm 0.04$ & $290 \pm 4$ & 
  $2.9 \pm 0.6$ & $1.0 \pm 0.5$ & $450 \pm 340$\\
  04 & $2.08 \pm 0.03$ & $190 \pm 10$ & $2.1 \pm 0.1$ & $300 \pm 10$ & 
  $5.5 \pm 0.1$ & $0.45 \pm 0.03$ & $280 \pm 20$ \\
  05 & $0.90 \pm 0.01$ & $170 \pm 3$ & $0.9 \pm 0.1$ & $280 \pm 10$ & 
  $2.8 \pm 0.1$ & $1.10 \pm 0.04$ & $420 \pm 20$\\
  06 & $0.91 \pm 0.03$ & $160 \pm 10$ & $0.76 \pm 0.03$ & $300 \pm 10$
  & 
  $3.3 \pm 0.2$ & $1.2 \pm 0.1$ & $410 \pm 40$ \\
  07 & $0.71 \pm 0.01$ & $270 \pm 10$ & $0.68 \pm 0.03$ & $450 \pm 10$
  & 
  $2.00 \pm 0.01$ & $0.90 \pm 0.02$ & $500 \pm 10$ \\
  08 & $1.0 \pm 0.1$ & $290 \pm 10$ & $0.64 \pm 0.02$ & $390 \pm 10$ & 
  $1.85 \pm 0.02$ & $0.72 \pm 0.03$ & $340 \pm 20$ \\
  09-1 & $1.05 \pm 0.01$ & $170 \pm 2$ & $1.0 \pm 0.1$ & $280 \pm 2$ & 
  $2.90 \pm 0.02$ & $0.72 \pm 0.02$ & $244 \pm 5$ \\
  09-2 & $0.37 \pm 0.01$ & $290 \pm 10$ & $0.36 \pm 0.03$ & $490 \pm
  20$ & 
  $0.07 \pm 0.01$ & $1.1 \pm 0.1$ & $740 \pm 80$ \\
  09-3 & $0.31 \pm 0.01$ & $370 \pm 20$ & $0.31 \pm 0.02$ & $580 \pm
  20$ & 
  $0.20 \pm 0.02$ & $0.4 \pm 0.1$ & $320 \pm 20$ \\
  09-4 & $0.66 \pm 0.03$ & $140 \pm 6$ & $0.7 \pm 0.1$ & $240 \pm 10$
  & 
  $0.07 \pm 0.01$ & $1.50 \pm 0.08$ & $540 \pm 60$ \\
  10-1 & $0.63 \pm 0.01$ & $240 \pm 6$ & $0.62 \pm 0.01$ & $380 \pm 2$
  & 
  $1.89 \pm 0.01$ & $0.77 \pm 0.01$ & $349 \pm 5$ \\
  10-2 & $0.45 \pm 0.01$ & $410 \pm 50$ & $0.5 \pm 0.1$ & $660 \pm 70$
  & 
  $0.24 \pm 0.05$ & $0.5 \pm 0.1$ & $410 \pm 60$ \\
  10-3 & $0.172 \pm 0.007$ & $780 \pm 100$ & $0.17 \pm 0.03$ & $1,200
  \pm 130$ & 
  $0.10 \pm 0.02$ & $0.4 \pm 0.1$ & $640 \pm 70$ \\
  10-4 & $0.21 \pm 0.01$ & $710 \pm 100$ & $0.22 \pm 0.05$ & $1,100
  \pm 140$ & 
  $0.13 \pm 0.03$ & $0.4 \pm 0.1$ & $660 \pm 100$ \\
  11 & $0.317 \pm 0.003$ & $440 \pm 10$ & $0.313 \pm 0.006$ & $710 \pm
  10$ & 
  $0.860 \pm 0.004$ & $0.6\pm 0.1$& $520 \pm 10$ \\
  12 & $0.521 \pm 0.002$ & $325 \pm 4$ & $0.51 \pm 0.01$ & $530 \pm 8$
  & 
  $1.46 \pm 0.01$ & $0.75 \pm 0.02$ & $500 \pm 10$ \\
  13 & \multicolumn{7}{c}{no central object} \\
  14 & \multicolumn{7}{c}{no central object} \\
  15 & $0.202 \pm 0.004$ & $270 \pm 15$ & -- & -- &
  $0.67 \pm 0.01$ & $1.60 \pm 0.04$ & $1,550 \pm 120$ \\
  16-1 & $0.25 \pm 0.01$ & $430 \pm 29$ & $0.2 \pm 0.03$ & $720 \pm
  50$ & 
  $0.02 \pm 0.01$ & $1.4 \pm 0.2$ & $1,750 \pm 490$ \\
  16-2 & $1.28 \pm 0.03$ & $190 \pm 10$ & $1.3 \pm 0.1$ & $300 \pm 20$
  & 
  $0.47 \pm 0.07$ & $0.66 \pm 0.09$ & $250 \pm 30$ \\
  17 & $0.026 \pm 0.001$ & $1,000 \pm 70$ & $0.025 \pm 0.003$ & $1,663
  \pm 107$ & 
  $0.0038 \pm 0.0008$ & $1.2 \pm 0.1$ & $2,980 \pm 610$ \\
  18 & $0.191 \pm 0.003$ & $450 \pm 20$ & $0.19 \pm 0.01$ & $720 \pm
  20$ & 
  $0.52 \pm 0.01$& $0.61 \pm 0.05$ & $560 \pm 40$ \\
  19 & $0.280 \pm 0.003$ & $410 \pm 10$ & $0.0211 \pm 0.0001$ & $620
  \pm 9$ & 
  $0.096 \pm 0.004$ & $0.66 \pm 0.03$ & $490 \pm 10$ \\
  20-1 & $0.50 \pm 0.01$ & $360 \pm 10$ & $0.48 \pm 0.02$ & $570 \pm
  10$ & 
  $1.35 \pm 0.02$ & $0.55 \pm 0.03$ & $400 \pm 20$ \\
  20-2 & $3.1 \pm 0.2$ & $40 \pm 3$ & $2.8 \pm 0.3$ & $70 \pm 10$ & 
  $0.05 \pm 0.04$& $2.0 \pm 0.5$ & $1,110 \pm 60$ \\
  20-3 & $0.42 \pm 0.01$ & $460 \pm 30$ & $0.42 \pm 0.04$ & $720 \pm
  40$ & 
  $0.27 \pm 0.02$ & $0.37 \pm 0.04$ & $400 \pm 20$ \\
  20-4 & $0.45 \pm 0.03$ & $360 \pm 40$ & $0.5 \pm 0.1$ & $530 \pm 40$
  & 
  $0.19 \pm 0.04$ & $0.5 \pm 0.1$ & $440 \pm 60$ \\
  21 & $0.322 \pm 0.002$ & $390 \pm 8$ & $0.31 \pm 0.01$ & $630 \pm 8$
  & 
  $0.872 \pm 0.007$ & $0.59 \pm 0.02$ & $470 \pm 10$ \\
  22 & $0.60 \pm 0.01$ & $280 \pm 10$ & $0.05 \pm 0.04$ & $480 \pm 20$
  & 
  $1.85 \pm 0.02$ & $1.06 \pm 0.03$ & $680 \pm 40$ \\
  23 & \multicolumn{7}{c}{no central object} \\
  24 & \multicolumn{7}{c}{no central object} \\
  25 & $1.080 \pm 0.005$ & $210 \pm 3$ & $1.1 \pm 0.4$ & $350 \pm 10$
  & 
  $3.1 \pm 0.1$ & $0.82 \pm 0.04$ & $360 \pm 20$ \\
  26 & $0.233 \pm 0.003$ & $300 \pm 10$ & $0.23 \pm 0.02$ & $510 \pm
  20$ & 
  $0.74 \pm 0.01$ & $1.23 \pm 0.04$ & $940 \pm 60$ \\
  27-1 & -- & -- & $0.32 \pm 0.03$& $870 \pm 50$ &
  $0.16 \pm 0.02$ & $0.37 \pm 0.05$ & $490 \pm 30$ \\
  27-2 & --& -- & $0.5 \pm 0.2$ & $300 \pm 60$ &
  $0.25 \pm 0.07$ & $0.7 \pm 0.1$ & $350 \pm 80$ \\
  28 & $0.088 \pm 0.001$ & $380 \pm 20$ & $0.083 \pm 0.007$ & $680 \pm
  30$ & 
  $0.27 \pm 0.01$ & $1.35 \pm 0.07$ & $1,500 \pm 170$ \\
  29 & $0.120 \pm 0.001$ & $650 \pm 20$ & $0.12 \pm 0.01$ & $1,030 \pm
  30$ & 
  $0.318 \pm 0.004$ & $0.53 \pm 0.04 $ & $720 \pm 40$ \\
  30 & $0.772 \pm 0.003$ & $140 \pm 1$ & $1.5 \pm 0.1$ & $230 \pm 5$ & 
  $4.64 \pm 0.08$ & $0.94 \pm 0.04$ & $270 \pm 20$ \\
  31 & $1.11 \pm 0.02$ & $270 \pm 10$ & -- & -- &
  $3.00 \pm 0.04$ & $0.59 \pm 0.04$ & $330 \pm 20$ \\
  32 & $0.550 \pm 0.002$ & $220 \pm 3$ & $0.52 \pm 0.09$ & $370 \pm 4$
  & 
  $1.53 \pm 0.01$ & $0.72 \pm 0.02$ & $382 \pm 9$ \\
  33 & $1.13 \pm 0.01$ & $150 \pm 2$ & $1.1\ pm 0.2$ & $250 \pm 3$ & 
  $3.19 \pm 0.02$ & $0.77 \pm 0.01$ & $234 \pm 4$ \\
  34-1 & $0.562 \pm 0.004$ & $300 \pm 10$ & $0.550 \pm 0.120$ & $475
  \pm 6$ & 
  $1.52 \pm 0.01$ & $0.59 \pm 0.02$ & $350 \pm 10$ \\
  34-2 & $5.85 \pm 0.05$ & $90 \pm 3$ & $5.78 \pm 0.32$ & $147 \pm 4$
  & 
  $1.8 \pm 0.1$ & $0.80 \pm 0.04$ & $140 \pm 10$ \\
  35 & $1.65 \pm 0.01$ & $140 \pm 2$ & -- & -- &
  $5.18 \pm 0.07$ & $1.05 \pm 0.03$ & $320 \pm 10$ \\
  36-1 & $0.473 \pm 0.003$ & $350 \pm 10$ & $0.45 \pm 0.02$ & $570 \pm
  10$ & 
  $1.33 \pm 0.01$ & $0.78 \pm 0.02$ & $550 \pm 10$ \\
  36-2 & $0.53 \pm 0.01$ & $90 \pm 3$ & $0.5 \pm 0.1$ & $460 \pm 46$ &
  $0.2 \pm 0.1$ & $0.8 \pm 0.2$ & $460 \pm 60$ \\
  37-1 & $1.12 \pm 0.01$ & $230 \pm 4$ & -- & -- & 
  $3.24 \pm 0.02$ & $0.65 \pm 0.02$ & $308 \pm 7$ \\
  37-2 & $1.00 \pm 0.01$ & $200 \pm 4$ & $0.98 \pm 0.05$ & $340 \pm
  10$ & 
  $0.24 \pm 0.01$ & $0.92 \pm 0.03$ & $390 \pm 20$ \\
  38 & $0.246 \pm 0.001$ & $410 \pm 10$ & $0.24 \pm 0.01$ & $670 \pm
  10$ & 
  $0.68 \pm 0.01$ & $0.71 \pm 0.02$ & $600 \pm 20$ \\
  39 & $0.61 \pm 0.01$ & $190 \pm 10$ & $0.57 \pm 0.03$ & $320 \pm 10$
  & 
  $1.68 \pm 0.05$ & $0.71 \pm 0.07$ & $280 \pm 30$ \\
  40 & $0.088 \pm 0.001$ & $860 \pm 25$ & $0.087 \pm 0.003$ & $1,350
  \pm 30$ & 
  $0.240 \pm 0.002$ & $0.49 \pm 0.02$ & $850 \pm 30$ \\
  41 & $0.65 \pm 0.02$ & $270 \pm 20$ & $0.64 \pm 0.07$ & $430 \pm 30$
  & 
  $1.81 \pm 0.06$ & $0.7 \pm 0.1$ & $400 \pm 60$ \\
  42 & $0.48 \pm 0.01$ & $410 \pm 16$ & $0.47 \pm 0.02$ &$660 \pm 20$
  & 
  $1.27 \pm 0.01$ & $0.42 \pm 0.01$ & $393 \pm 6$ \\
  43 & $2.00 \pm 0.02$ & $200 \pm 5$ & $1.94 \pm 0.03$ & $320 \pm 4$ & 
  $5.42 \pm 0.03$ & $0.62 \pm 0.01$ & $246 \pm 4$\\
  44 & $0.453 \pm 0.004$ & $400 \pm 10$ & $0.44 \pm 0.01$ & $650 \pm
  8$ & 
  $1.22 \pm 0.01$ & $0.57 \pm 0.01$ & $475 \pm 8$ \\
  45 & $1.34 \pm 0.01$ & $240 \pm 6$ & $1.32 \pm 0.04$ & $390 \pm 6$ & 
  $3.65 \pm 0.02$ & $0.63 \pm 0.02$ & $307 \pm 7$ \\
  46 & $0.541 \pm 0.002$ & $304 \pm 4$ & $0.53 \pm 0.01$ & $500 \pm 6$
  & 
  $1.49 \pm 0.01$ & $0.70 \pm 0.02$ & $440 \pm 10$ \\
  47 & $1.9 \pm 0.1$ & $170 \pm 15$ & $1.9 \pm 0.3$ & $270 \pm 23$ & 
  $5.02 \pm 0.25$ & $0.5 \pm 0.1$ & $180 \pm 30$ \\
  48 & $0.430 \pm 0.001$ & $310 \pm 3$ & $0.4 \pm 0.1$ & $500 \pm 6$ & 
  $1.19 \pm 0.01$ & $0.72 \pm 0.02$ & $450 \pm 20$ \\
  49 & $0.50 \pm 0.01$ & $380 \pm 10$ & $0.49 \pm 0.02$ & $630 \pm 20$
  & 
  $1.40 \pm 0.01$ & $0.75 \pm 0.03$ & $570 \pm 20$ \\
  50 & $0.365 \pm 0.002$ & $420 \pm 7$ & $0.350 \pm 0.001$ & $680 \pm
  10$ & 
  $1.10 \pm 0.01$ & $0.73 \pm 0.02$ & $620 \pm 20$ \\
  51 & $0.676 \pm 0.004$ & $330 \pm 6$& -- & -- &
  $1.84 \pm 0.01$ & $0.65 \pm 0.01$ & $432 \pm 7$ \\
  52 & $0.282 \pm 0.002$ & $500 \pm 12$ & $0.277 \pm 0.004$ & $810 \pm
  10$ & 
  $0.76 \pm 0.01$ & $0.55 \pm 0.02$ & $570 \pm 20$ \\
  53 & $0.7 \pm 0.1$ & $300 \pm 10$ & $0.64 \pm 0.05$ & $490 \pm 20$ & 
  $1.79 \pm 0.04$ & $0.68 \pm 0.07$ & $415 \pm 40$ \\
  \hline
\end{longtable}

\begin{longtable}{lcclc}
  \caption{The morphological parameters for the final objects of each
    simulation.  The first column states the name of the simulation
    according to Tab.~\ref{tab:init-cond}.  The second column gives the
    clumpiness $C$ followed by the ellipticity $e$ (measured at the
    half-mass radius) of the object.  In the last column we state the
    Fourier parameter $A4$, which describes disky and boxy deviations
    of the shape of our objects.  Again we state the value measured at
    the half-mass radius.} \\
  \hline
  \label{tab:shape}
  Simulation & $C$ & $e$ & \multicolumn{1}{c}{$A_{4}$} & Surviving
  SC(s) \\ 
  \hline
  \endfirsthead
  \hline
  Simulation & $C$ & $e$ & \multicolumn{1}{c}{$A_{4}$} & Surviving
  SC(s) \\ 
  \hline
  \endhead
  N015M107RH025RS025S30 & $0.04$ & $0.06$ & $+0.004$ & $0$ \\
  P015M107RH025RS025S30 & $0.05$ & $0.11$ & $-0.015$ & $0$ \\
  N030M107RH025RS025S30 & $0.04$ & $0.22$ & $-0.04$ & $0$ \\
  P030M107RH025RS025S30 & $0.04$ & $0.21$ & $-0.02$ & $0$ \\
  N015M107RH050RS025S30 & $0.04$ & $0.27$ & $+0.01$ & $0$ \\
  P015M107RH050RS025S30 & $0.04$ & $0.27$ & $-0.02$ & $0$ \\
  N030M107RH050RS025S30 & $0.03$ & $0.16$ & $-0.03$ & $0$ \\
  P030M107RH050RS025S30 & $0.03$ & $0.20$ & $-0.02$ & $0$ \\
  N015M107RH100RS025S30-1 & $0.05$ & $0.18$ & $-0.024$ & $0$ \\
  N015M107RH100RS025S30-2 & $0.10$ & $0.21$ & $+0.19$ & $1$ \\
  N015M107RH100RS025S30-3 & $0.11$ & $0.44$ & $-0.05$ & $1$ \\
  N015M107RH100RS025S30-4 & $0.10$ & $0.33$ & $-0.01$ & $2$ \\
  P015M107RH100RS025S30-1 & $0.07$ & $0.39$ & $+0.03$ & $2$ \\
  P015M107RH100RS025S30-2 & $0.19$ & $0.19$ & $+0.01$ & $7$ \\
  P015M107RH100RS025S30-3 & $0.26$ & $0.78$ & $-0.034$ & $6$ \\
  P015M107RH100RS025S30-4 & $0.10$ & $0.33$ & $-0.01$ & $2$ \\
  N015M107RH100RS025S15 & $0.04$ & $0.47$ & $-0.05$ & $0$ \\
  P015M107RH100RS025S15 & $0.05$ & $0.54$ & $-0.12$ & $0$\\
  N015M107RH100RS025S60 & \multicolumn{3}{c}{no central object} & 15 \\ 
  P015M107RH100RS025S60 & \multicolumn{3}{c}{no central object} & 15 \\ 
  N015M107RH100RS050S30 & $0.06$ & $0.37$ & $-0.106$ & $0$ \\
  P015M107RH100RS050S30-1 & $0.08$ & $0.11$ & $+0.08$ & $0$ \\
  P015M107RH100RS050S30-2 & $0.04$ & $0.29$ & $0.02$ & $6$ \\
  N015M107RH100RS100S30 & $0.26$ & $0.80$ & $+0.32$ & $0$ \\
  P015M107RH100RS100S30 & $0.12$ & $0.12$ & $+0.28$ & $0$ \\
  N030M107RH100RS025S30-1 & $0.04$ & $0.28$ & $-0.004$ & $0$ \\
  N030M107RH100RS025S30-2 & $0.03$ & $0.11$ & $-0.014$ & $0$ \\
  N030M107RH100RS025S30-3 & $0.09$ & $0.69$ & $+0.15$ & $0$ \\
  N030M107RH100RS025S30-4 & $0.04$ & $0.13$ & $-0.07$ & $0$ \\
  P030M107RH100RS025S30-1 & $0.05$ & $0.03$ & $+0.02$ & $0$ \\
  P030M107RH100RS025S30-2 & $0.12$ & $0.33$ & $+0.021$ & $7$ \\
  P030M107RH100RS025S30-3 & $0.19$ & $0.22$ & $-0.019$ & $6$ \\
  P030M107RH100RS025S30-4 & $0.15$ & $0.26$ & $-0.013$ & $6$ \\
  N030M107RH100RS025S15 & $0.05$ & $0.03$ & $-0.02$ & $0$ \\
  P030M107RH100RS025S15 & $0.09$ & $0.17$ & $+0.18$ & $0$ \\
  N030M107RH100RS025S60 & \multicolumn{3}{c}{no central object} & 30 \\ 
  P030M107RH100RS025S60 & \multicolumn{3}{c}{no central object} & 30 \\ 
  N060M107RH100RS025S30 & $0.02$ & $0.15$ & $+0.01$ & $0$ \\
  N030M107RH100RS050S30 & $0.07$ & $0.52$ & $+0.07$ & $0$ \\
  P030M107RH100RS050S30-1 & $0.12$ & $0.20$ & $+0.0274$ & $1$ \\
  P030M107RH100RS050S30-2 & $0.16$ & $0.27$ & $+0.001$ & $6$ \\
  N030M107RH100RS100S30 & $0.08$ & $0.19$ & $-0.004$ & $0$ \\
  P030M107RH100RS100S30 & $0.07$ & $0.41$ & $+0.523$ & $0$\\
  N015M407RH025RS025S30 & $0.04$ & $0.48$ & $-0.02$ & $0$ \\
  P015M407RH025RS025S30 & $0.11$ & $0.13$ & $-0.12$ & $0$ \\
  N030M407RH025RS025S30 & $0.04$ & $0.10$ & $-0.03$ & $0$ \\
  P030M407RH025RS025S30 & $0.05$ & $0.26$ & $+0.08$ & $0$ \\
  N015M407RH050RS025S30-1 & $0.04$ & $0.27$ & $+0.064$ & $0$ \\
  N015M407RH050RS025S30-2 & $0.02$ & $0.09$ & $+0.01$ & $0$ \\
  P015M407RH050RS025S30-1 & $0.07$ & $0.49$ & $-0.13$ & $0$ \\
  P015M407RH050RS025S30-2 & $0.07$ & $0.36$ & $+0.11$ & $2$ \\
  N030M407RH050RS025S30-1 & $0.03$ & $0.40$ & $+0.06$ & $0$ \\
  N030M407RH050RS025S30-2 & $0.03$ & $0.18$ & $+0.01$ & $0$ \\
  P030M407RH050RS025S30 & $0.05$ & $0.60$ & $-0.04$ & $0$ \\
  N015M407RH100RS025S30 & $0.04$ & $0.03$ & $-0.09$ & $0$ \\
  P015M407RH100RS025S30-1 & $0.09$ & $0.01$ & $-0.2$ & $1$ \\
  P015M407RH100RS025S30-2 & $0.24$ & $0.13$ & $-0.017$ & $4$ \\
  P015M407RH100RS025S30-3 & $0.27$ & $0.66$ & $+0.05$ & $2$ \\
  N030M407RH100RS025S30 & $0.04$ & $0.12$ & $+0.04$ & $0$ \\
  P030M407RH100RS025S30 & $0.08$ & $0.11$ & $+0.08$ & $1$ \\
  N015M108RH025RS025S30 & $0.03$ & $0.14$ & $-0.02$ & $0$ \\
  P015M108RH025RS025S30 & $0.07$ & $0.23$ & $+0.007$ & $0$ \\
  N030M108RH025RS025S30 & $0.04$ & $0.32$ & $-0.09$ & $0$ \\
  P030M108RH025RS025S30 & $0.05$ & $0.10$ & $-0.01$ & $0$ \\
  N015M108RH050RS025S30 & $0.07$ & $0.19$ & $+0.01$ & $0$ \\
  P015M108RH050RS025S30 & $0.07$ & $0.23$ & $+0.007$ & $0$ \\
  N030M108RH050RS025S30 & $0.07$ & $0.20$ & $+0.18$ & $0$ \\
  P030M108RH050RS025S30 & $0.03$ & $0.10$ & $+0.041$ & $0$ \\
  N015M108RH100RS025S30 & $0.07$ & $0.20$ & $+0.181$ & $0$ \\
  P015M108RH100RS025S30 & $0.03$ & $0.08$ & $-0.015$ & $0$ \\
  N030M108RH100RS025S30 & $0.08$ & $0.37$ & $-0.14$ & $0$ \\
  P030M108RH100RS025S30 & $0.03$ & $0.10$ & $+0.04$ & $0$ \\
  \hline
\end{longtable}

\begin{longtable}{lrcrl} 
  \caption{Table of all results regarding the velocity dispersion and
    $\delta$ parameter.  The first column gives the name of the
    simulation, the second column gives the central velocity
    dispersion $\sigma_{\rm 0,mean}$ measured in a circular area of $r
    = 10$~pc.  We give the mean value out of the three projections
    along the coordinate axes.  With $\sigma_{\rm 0,fit}$ we fit a
    Plummer profile to the radial velocity dispersion profil and give
    the central value of the fitting curve.  $\sigma_{500}$ is the
    overall velocity dispersion measured using all stars within a
    radius of $500$~pc.  Finally, the last column $\delta_{500}$
    denotes the maximum deviation from the mean radial velocity found
    in a 25 by 25~pc pixel within an area of $500$~pc radius.} \\
  \hline
  \label{tab:vel1}
  Simulation & $\sigma_{\rm 0,mean}$ & $\sigma_{\rm 0,fit}$ &
  $\sigma_{500}$ & $\delta_{500}$ \\
  & $[km/s]$ & $[km/s]$ & $[km/s]$ & \\
  \hline
  \endfirsthead
  \hline
  Simulation & $\sigma_{\rm 0,mean}$ & $\sigma_{\rm 0,fit}$ &
  $\sigma_{500}$ & $\delta_{500}$ \\
  & $[km/s]$ & $[km/s]$ & $[km/s]$ & \\
  \hline
  \endhead
  N015M107RH025RS025S30 & $5.6$ & $5.63 \pm 0.04$ & $5.0$ & $0.89$ \\
  P015M107RH025RS025S30 & $5.3$ & $5.4 \pm 0.1$ & $4.8$ & $1.35$ \\
  N030M107RH025RS025S30 & $5.9$ & $6.32 \pm 0.08$ & $5.5$ & $0.79$ \\
  P030M107RH025RS025S30 & $5.5$ & $5.70 \pm 0.03$ & $5.3$ & $0.89$ \\
  N015M107RH050RS025S30 & $6.6$ & $7.52 \pm 0.05$ & $5.7$ & $0.62$ \\
  P015M107RH050RS025S30 & $4.7$ & $4.7 \pm 0.1$ & $4.4$ & $0.85$ \\
  N030M107RH050RS025S30 & $6.8$ & $7.52 \pm 0.05$ & $6.9$ & $0.48$ \\
  P030M107RH050RS025S30 & $5.6$ & $6.07 \pm 0.04$ & $5.0$ & $1.03$ \\
  N015M107RH100RS025S30-1 & $5.5$ & $5.50 \pm 0.05$ & $5.6$ & $ 1.15$
  \\ 
  N015M107RH100RS025S30-2 & $7.6$ & $8.40 \pm 0.05$ & $8.1$ & $1.52$
  \\ 
  N015M107RH100RS025S30-3 & $7.8$ & $8.40 \pm 0.08$ & $8.5$ & $1.65$
  \\ 
  N015M107RH100RS025S30-4 & $7.9$ & $8.11 \pm 0.02$ & $8.3$ & $1.73$
  \\ 
  P015M107RH100RS025S30-1 & $8.5$ & $9.70 \pm 0.07$ & $8.1$ & $ 0.51$
  \\ 
  P015M107RH100RS025S30-2 & $9.1$ & $10.0 \pm 0.1$ & $8.2$ & $1.50$ \\ 
  P015M107RH100RS025S30-3 & $6.3$ & $6.3 \pm 0.3$ & $8.5$ & $1.70$ \\
  P015M107RH100RS025S30-4 & $7.9$ & $8.10 \pm 0.02$ & $8.3$ & $1.70$
  \\ 
  N015M107RH100RS025S15 & $7.0$ & $7.22 \pm 0.02$ & $7.1$ & $0.69$ \\
  P015M107RH100RS025S15 & $7.6$ & $7.71 \pm 0.03$ & $7.5$ & $0.71$ \\
  N015M107RH100RS025S60 & \multicolumn{4}{c}{no central object} \\
  P015M107RH100RS025S60 & \multicolumn{4}{c}{no central object} \\
  N015M107RH100RS050S30 & $9.2$ & $9.82 \pm 0.03$ & $8.3$ & $1.01$ \\ 
  P015M107RH100RS050S30-1 & $7.3$ & $7.95 \pm 0.04$ & $8.0$ & $1.6$ \\
  P015M107RH100RS050S30-2 & $8.0$ & $3.0 \pm 0.2$ & $8.6$ & $2.9$ \\
  N015M107RH100RS100S30 & $5.7$ & $6.0 \pm 0.1$ & $7.8$ & $1.29$ \\
  P015M107RH100RS100S30 & $8.2$ & $8.1 \pm 0.1$ & $7.4$ & $1.1$ \\
  N030M107RH100RS025S30-1 & $7.5$ & $9.95 \pm 0.07$ & $9.0$ & $0.68$
  \\ 
  N030M107RH100RS025S30-2 & $11.3$ & $11.8 \pm 0.20$ & $14.5$ & $0.2$
  \\ 
  N030M107RH100RS025S30-3 & $8.4$ & $8.95 \pm 0.03$ & $7.0$ & $0.5$ \\
  N030M107RH100RS025S30-4 & $7.3$ & $7.87 \pm 0.05$ & $8.3$ & $1.7$ \\
  P030M107RH100RS025S30-1 & $8.5$ & $6.1 \pm 0.1$ & $7.7$ & $1.15$ \\
  P030M107RH100RS025S30-2 & $6.1$ & $6.1 \pm 0.1$ & $6.3$ & $0.9$ \\
  P030M107RH100RS025S30-3 & $7.5 $ & $7.89 \pm 0.06$ & $6.8$ & $1.2$
  \\ 
  P030M107RH100RS025S30-4 & $6.4$ & $6.4 \pm 0.3$ & $7.3$ & $1.3$ \\  
  N030M107RH100RS025S15 & $7.2$ & $7.57 \pm 0.02$ & $7.2$ & $0.45$ \\
  P030M107RH100RS025S15 & $8.1$ & $8.64 \pm 0.01$ & $6.9$ & $0.76$ \\
  N030M107RH100RS025S60 & \multicolumn{4}{c}{no central object} \\
  P030M107RH100RS025S60 & \multicolumn{4}{c}{no central object} \\
  N060M107RH100RS025S30 & $7.4$ & $7.78 \pm 0.05$ & $7.2$ & $0.49$ \\
  N030M107RH100RS050S30 & $9.0$ & $9.46\pm 0.04 $ & $8.0$ & $1.16$ \\ 
  P030M107RH100RS050S30-1 & $7.3$ & $7.25 \pm 0.06$ & $7.3$ & $0.9$ \\ 
  P030M107RH100RS050S30-2 & $9.0$ & $8.8 \pm 0.1$ & $8.0$ & $0.9$ \\
  N030M107RH100RS100S30 & $7.5$ & $7.49 \pm 0.06$ & $8.0$ & $0.73$ \\
  P030M107RH100RS100S30 & $8.5$ & $8.47 \pm 0.05$ & $8.6$ & $1.1$ \\
  N015M407RH025RS025S30 & $11.5$ & $12.01 \pm 0.04$ & $10.8$ & $0.86$
  \\ 
  P015M407RH025RS025S30 & $11.9$ & $12.6 \pm 0.2$ & $11.5$ & $1.21$ \\
  N030M407RH025RS025S30 & $11.9$ & $12.72 \pm 0.07$ & $11.0$ & $0.92$
  \\ 
  P030M407RH025RS025S30 & $10.4$ & $10.44 \pm 0.05$ & $11.0$ & $1.15$ \\
  N015M407RH050RS025S30-1 & $13.5$ & $14.31 \pm 0.06$ & $12.7$ & $0.72$
  \\
  N015M407RH050RS025S30-2 & $9.1$ & $9.58 \pm 0.04$ & $8.9$ & $0.5$ \\
  P015M407RH050RS025S30-1 & $11.7$ & $13.1 \pm 0.2$ & $11.3$ & $1.02$ \\
  P015M407RH050RS025S30-2 & $11.5$ & $11.9 \pm 0.3$ & $11.2$ & $0.6$ \\
  N030M407RH050RS025S30-1 & $15.4$ & $17.75 \pm 0.04$ & $12.4$ & $0.54$
  \\
  N030M407RH050RS025S30-2 & $15.0$ & $15.8 \pm 0.1$ & $12.0$ & $0.5$
  \\ 
  P030M407RH050RS025S30 & $11.2$ & $11.98 \pm 0.08$ & $11.0$ & $0.89$
  \\ 
  N015M407RH100RS025S30 & $16.8$ & $17.7 \pm 0.1$ & $16.3$ & $0.59$ \\ 
  P015M407RH100RS025S30-1 & $13.4$ & $13.8 \pm 0.1$ & $13.5$ & $0.61$
  \\ 
  P015M407RH100RS025S30-2 & $23.0$ & $18.12 \pm 0.09$ & $15.0$ & $1.1$
  \\
  P015M407RH100RS025S30-3 & $22.3$ & $12.0 \pm 0.2$ & $16.0$ & $0.79$
  \\ 
  N030M407RH100RS025S30 & $12.3$ & $12.67 \pm 0.05$ & $13.8$ & $0.51$
  \\ 
  P030M407RH100RS025S30 & $15.7$ & $15.4 \pm 0.1$ & $14.3$ & $0.69$ \\
  N015M108RH025RS025S30 & $ 18.0$ & $18.1 \pm 0.1$ & $18.1$ & $0.55$
  \\ 
  P015M108RH025RS025S30 & $18.4$ & $21.0 \pm 0.1$ & $17.1$ & $1.01$ \\ 
  N030M108RH025RS025S30 & $15.7$ & $15.65 \pm 0.08$ & $16.4$ & $1.17$
  \\ 
  P030M108RH025RS025S30 & $16.5$ & $16.9 \pm 0.1$ & $16.4 $ & $0.96$
  \\ 
  N015M108RH050RS025S30 & $23.9$ & $26.95 \pm 0.07$ &$20.0$ & $0.60$
  \\ 
  P015M108RH050RS025S30 & $15.5$ & $15.9 \pm 0.1$ & $15.4$ & $1.01$ \\ 
  N030M108RH050RS025S30 & $22.4$ & $23.33 \pm 0.07$ & $21.3$ & $0.47$
  \\ 
  P030M108RH050RS025S30 & $16.0$ & $16.20 \pm 0.03$ & $16.9$ & $1.09$
  \\ 
  N015M108RH100RS025S30 & $30.4$ & $32.51 \pm 0.09$ & $26.8$ & $0.77$
  \\ 
  P015M108RH100RS025S30 & $24.8$ & $27.55 \pm 0.08$ & $22.0$ & $0.66$
  \\ 
  N030M108RH100RS025S30 & $30.9$ & $32.64 \pm 0.06$ & $28.3$ & $0.63$
  \\ 
  P030M108RH100RS025S30 & $26.2$ & $28.5 \pm 0.1$ & $23.1$ & $0.67$
  \\ 
  \hline
\end{longtable}
\end{center}

\label{lastpage}

\end{document}